\documentclass[hidelinks,aps,pra,twocolumn,superscriptaddress,10pt]{revtex4-2} 
\usepackage{amsmath}
\usepackage{amssymb}
\usepackage{graphicx}
\usepackage{color}
\usepackage{url}
\usepackage{listings}
\newcommand{\be}[1]{\begin{equation}\label{#1}}
\newcommand{\ee}{\end{equation}}
\newcommand{\bea}[1]{\begin{eqnarray}\label{#1}}
\newcommand{\eea}{\end{eqnarray}}
\newcommand{\no}{\nonumber \\}
\newcommand{\Fig}[1]{Fig.(\ref{#1})}

\newcommand{\Eq}[1]{Eq.(\ref{#1})}
\newcommand{\App}[1]{Appendix~\ref{#1}}
\newcommand{\Sec}[1]{Section~\ref{#1}}
\newcommand{\bsub}{\begin{subequations}}
\newcommand{\esub}{\end{subequations}}
\newcommand{\bwt}{\begin{widetext}}
\newcommand{\ewt}{\end{widetext}}

\usepackage{color}

\def\trm#1{\textrm{#1}}
\def\tit#1{\textit{#1}}

\def\defn{\overset{\trm{def}}{=}}

\def\a0{{\alpha_0}}

\def\da0{{\dot{\alpha}_0}}

\def\myoverDefn#1#2{\hbox{\space \raise-2mm\hbox{$\textstyle{#1} \atop \scriptstyle{#2}$} }}

\def\G{{\Gamma}}

\def\a{{\alpha}}

\def\dag{\dagger}

\def\Im{\textrm{Im}}

\def\rp{r_{P}}
\def\rp2{r_{p}^{2}}
\def\Tr{\textrm{Tr}}

\def\N{{\mathcal{N}}}
\def\tN{{\mathcal{\tilde{N}}}}
\newcommand{\half}{\frac{1}{2}}

\newcommand{\ket}[1]{|#1\rangle}
\newcommand{\bra}[1]{\langle #1|}

\newcommand{\IP}[2]{\langle {#1} | {#2} \rangle}  
\newcommand{\EV}[1]{\langle {#1} \rangle}          

\begin{document}

\title{Numerical evidence for a bipartite pure state entanglement witness\\ 
from approximate analytical diagonalization}
\author{Paul M. Alsing}\email{corresponding author: palsing@albany.edu}
\affiliation{University at Albany-SUNY, Albany, NY 12222, USA}
\author{Richard J. Birrittella}
\affiliation{Booz Allen Hamilton, 99 Otis St, Rome, NY, 13411, USA}
\affiliation{Air Force Research Laboratory, Information Directorate, 525 Brooks Rd, Rome, NY, 13411, USA}

\date{\today}

\begin{abstract}
We show numerical evidence for a bipartite $d\times d$ pure state entanglement witness that is readily calculated from the wavefunction coefficients directly, without the need for the numerical computation of eigenvalues. This is accomplished by using an approximate analytic diagonalization of the bipartite state that captures dominant contributions to the negativity of the partially transposed state.
We relate this entanglement witness to the Log Negativity, and show that it exactly agrees with it for the class of pure states whose quantum amplitudes form a positive  Hermitian matrix. 
In this case, the Log Negativity is given by the negative logarithm of the purity of the amplitudes consider as a density matrix. 
In other cases, the witness forms a lower bound to the exact, numerically computed Log Negativity. 
The formula for the approximate Log Negativity achieves equality 
with the exact Log Negativity for the case of an arbitrary pure state of two qubits, which we show analytically.
We compare these results to a witness of entanglement given by the linear entropy.
Finally, we explore an attempt to extend these pure state results to mixed states. We show that the Log Negativity for this approximate formula is exact  on the class of  pure state decompositions for which the quantum amplitudes of each pure state form a positive  Hermitian matrix. 
\end{abstract}

\maketitle

\section{Introduction}\label{sec:Intro}
Entanglement is a key quantum resource for information tasks  including timing/sensing \cite{Toth:2014}, communication and networking \cite{QWkshpReport:2014}, and computation \cite{NC:2000}. 
It is a fundamental quantum phenomena, first discussed by Schr\"{o}dinger 
\cite{Schrodinger:1935, Schrodinger:1936} in response to the criticisms of the incompleteness of quantum mechanics by Einstein, Podolsky and Rosen \cite{EPR:1935}, manifesting its effects both at the atomic scale, and possibly even the cosmic scale in theories of formation of spacetime itself \cite{Swingle:2017}.
One of the important research areas in modern quantum information science is the difficult issue of the quantification and characterization of entanglement from bipartite to multi-partite quantum systems
(see \cite{Horodecki4:2009,Zyczkowski_2ndEd:2020} and references therein).

For pure states, an important measure of bipartite entanglement is the 
von Neuman entropy (VNE) $S(\rho_a) = -\sum_n \lambda_n \log_2 \lambda_n$
of the reduced subsystem
$\rho_a=\Tr_b[\rho_{ab}]$  of the composite state $\rho_{ab}$. 
This, of course, requires the diagonalization of $\rho_a$ to obtain its eigenvalues $\{\lambda_n\}$.
Another widely used and easy to compute measure of pure state bipartite entanglement is the linear entropy $LE = 1-\Tr[\rho^2_a]$, whose popularity stems from not having to compute the eigenvalues of $\rho_a$.

While the above measures of entanglement are applicable to pure states, entanglement measures for mixed states are harder to come by \cite{Horodecki4:2009}.  Measures of mixed state entanglement exists only for particular systems, such as Wotter's concurrence \cite{Wootters:1998} for a pair of qubits and for a qubit-qutrit system. A widely used ``measure" of entanglement for both pure and mixed states, both discrete and continuous variable, 
is the Log Negativity ($LN$) \cite{Peres:1996,Horodecki:1997,Agarwal:2013} given by
$LN(\rho_{ab}) = \log_2(1+2\N)$ where the negativity $\N$ is given by the sum of the absolute values of the negative eigenvalues of the partial transpose of  $\rho_{ab}$. This entanglement monotone is based on the fact a separable state
$\rho^{(sep)}_{ab}=\sum_i p_i \rho^{(i)}_a\otimes \rho^{(i)}_b$ remains positive under partial transpose.
It is not a true entanglement measure since it does not detect bound entanglement (i.e. states that are entangled, yet have positive partial transposes).

In this work, we present numerical evidence for a pure state entanglement witness that exactly agrees with the $LN$ when the quantum amplitudes $\{c_{nm}\}\to C$ of 
$\ket{\psi}_{ab}=\sum_{n=0}^{d-1}  \sum_{m=0}^{d-1} c_{nm}\ket{n,m}_{ab}$ 
(a $d^2\times 1$ column vector)
forms a
$d\times d$  positive Hermitian matrix, $C^\dag = C$, $C\ge 0$, with
normalization $\IP{\psi_{ab}}{\psi_{ab}}=\Tr[C\,C^\dag]=1$.
That is,
\bea{pure:state:summary} 
\trm{if}\;\; C &\defn& \frac{\rho}{\sqrt{\Tr[\rho^2]}},\quad  \trm{with}\;\; \rho=\rho^\dag,\no
\Rightarrow LN_e &=& LN_a = -\log_2(\Tr[\rho^2])\;\; \trm{for}\;\; \Tr[\rho]=1,
\eea
where $\Tr[\rho^2]$ is the purity of $\rho$.
Here we have denoted our approximate expression for the Log Negativity as $LN_{a}$ ($a$ for \tit{approximate}), and distinguish it from the exact expression for the Log Negativity, $LN_{e}$ ($e$ for \tit{exact}) obtained by numerical computation of the eigenvalues of the partial transpose of the pure state density matrix $\rho_{ab} \defn \ket{\psi}_{ab}\bra{\psi}$).
For the general case, when $C$ is an arbitrary complex matrix,  
this witness acts as a lower bound to the exact Log Negativity,
$LN_{a}(\rho_{ab})\le LN_{e}(\rho_{ab})$.
For arbitrary complex $C$, equality is reached  only for the case of two qubits, which we derive analytically.
We compare our witness with that of the $LE$.

While we are not able to provide a bona fide  analytic deriation of $LN_{a}$, we do provide a plausibility argument modeled on the analytic diagonalization of correlated pure states of the form 
$\ket{\psi}_{ab} = \sum_{n=0}^{d-1} c_n \ket{n,n}$ by Agarwal \cite{Agarwal:2013}. 
This form encompasses both discrete and continuous variable states (where the latter is truncated in each subspace to a maximum Fock number state $\ket{M}$ with $d=M+1$  
such that  $|c_{n>M}|^2<\epsilon\ll 1$ for some arbitrary chosen $\epsilon$).
Agarwal's derivation can be interpreted as a generalization of the analytic diagonalization of the Schmidt decomposition of the pure state $\ket{\psi}_{ab}$, which yields $LN_{e}$ in terms of the magnitudes of the \tit{complex} quantum amplitudes $\{c_n\}$.

We subsequently attempt to extend our witness from pure states to mixed states, with limited  success, though with interesting special cases, in the sense that we 
are able to provided numerical evidences that 
\be{mixed:state:summary}
\mathcal{LN}_a^{avg}(\rho) \leq LN_e(\rho) \leq LN_a(\rho)
\ee
when the mixed state is written as a pure state decomposition (PSD) 
$\rho_{ab} = \sum_i p_i\,\ket{\psi_i}_{ab}\bra{\psi_i}$ 
with the quantum amplitude matrix $C_i$
of each pure state component $\ket{\psi_i}_{ab} = \sum_{n m} C_{i,nm}\ket{n,m}_{ab}$  uniformly generated (over the Haar measure) as a positive Hermitian matrix, $C_i = C_i^\dag\ge0$.
%
%
In \Eq{mixed:state:summary} we have defined the  
\tit{average} of our approximate Log Negativity on a PSD as
\be{LN:avg:defn}
\mathcal{LN}_a^{avg}(\rho_{ab}) \defn  \sum_i p_i \,LN_a(\ket{\psi_i}_{ab}),
\ee
where the notation $LN_a(\ket{\psi_i}_{ab})$
 is used to denote our witness formula for pure states, defined in \Eq{LogNeg:Approx:line:2}.
 The rightmost term $LN_a(\rho)$ in \Eq{mixed:state:summary} is computed with our  generalized mixed state formula ansatz given in \Eq{Neg:LogNeg:Approx:rhoab:line:3}.
Lastly, we find that the second inequality is saturated, $LN_e(\rho) = LN_a(\rho)$ 
in \Eq{mixed:state:summary}, 
if the uniformly randomly generated $C_i$ matrices above are real, i.e. $C_i=C_i^T$.
(A discussion of the uniform generation of Hermitian matrices over the Haar measure is given in \App{app:Haar:purity}).


This paper is outlined as follows.
In \Sec{sec:Agaral:calc:p57-58} we discuss Agarwal's derivation \cite{Agarwal:2013} of the analytic diagonalization of correlated pure states of the form $\ket{\psi}_{ab} = \sum_n c_n \ket{n,n}$ and present an analytic formula for 
$LN_{e}$ in terms of the the Schmidt coefficients of the pure state.
In \Sec{sec:Ansatz:Witness} we present our ansatz for an entanglement witness 
$LN_{a}$ and numerical evidence for it on the pure state 
$\ket{\psi}_{ab} = \sum_{nm} c_{nm} \ket{n,m}$ based solely on its quantum amplitudes $\{c_{nm}\}$ without the need for numerical diagonalization of $\rho_{ab} = \ket{\psi}_{ab}\bra{\psi}$. We present numerical evidence that 
$LN_{a}\equiv LN_{e}$ when $\{c_{nm}\}\to C$ considered as a complex matrix is positive and Hermitian and that 
$LN_{a}=0$ on separable pure states. While not a formal proof, we present a plausibility argument for the derivation of $LN_{a}$ inspired by the previous Agarwal's derivation \cite{Agarwal:2013} in \Sec{sec:Agaral:calc:p57-58}, based on an ansatz for the dominant analytic contributions to the Negativity.
For $C$ an arbitrary complex matrix (subject to normalization), we present numerical evidence that 
$LN_{a}$ acts as a lower bound  to  $LN_{e}$ (better than $LE$) and hence acts as an entanglement witness. 
For the case of two arbitrary qubits, we analytically show that our approximate formula for the Log Negativity agrees with the exact formula.
In \Sec{sec:dms} we attempt to provide a generalization of $LN_{a}$ to mixed states, that reduces to the original formula for pure states, and again does not require numerical diagonalization of matrices derived from $\rho_{ab}$.  This generalization is zero on separable states 
$\rho_{ab} = \sum_i p_i\,\rho_a^{(i)}\otimes\rho_b^{(i)}$ only when either or both $\rho_a^{(i)}, \rho_b^{(i)}$ are real.
While this witness does detect a wide class of separable states, it does not detect them all (a notoriously difficult problem in its own right \cite{Horodecki4:2009, Zyczkowski_2ndEd:2020}). By examining Werner states of arbitrary dimension  
(mixing a generalized Bell state with a maximally mixed state of 
appropriate dimension), we are led to an ansatz for a mixed state witness for which we can numerically show that
\Eq{mixed:state:summary} holds over pure state decompositions (PSD)
for which the quantum amplitudes of the pure state components, when considered as $d \times d$ matrices, are positive and Hermitian.
In \Sec{sec:Conclusion}  we summarize our results and present our conclusions. 

The intent and focus of this work is on demonstrating that 
the information contained solely within the pure (mixed) state 
quantum amplitudes (matrix elements) directly provides inherent entanglement information that is
normally associated with the eigenvalues of matrices derived from the quantum state (e.g. partial transpose, reduced density matrices, etc.). While the numerical computation of eigenvalues does not present a practical impediment to the calculation of entanglement measures, it is the surprising relationship 
(and unexpected  equality for certain classes of physically relevant pure and mixed states) of the proposed  entanglement witness $LN_a$ to the exact (numerically computed) Log Negativity $LN_e$, that was the impetus for this current investigation.

\section{The Log Negativity for arbitrary pure states}\label{sec:Agaral:calc:p57-58}
Any bipartite pure state $\ket{\psi}_{ab}~\equiv~\sum_{n=0}^{M}\sum_{m=0}^{M}c_{nm}\ket{nm}_{ab}$ with complex amplitudes $c_{nm}\in\mathbb{C}$ and computational dual orthonormal  basis states $\ket{n m}_{ab}$ admits a Schmidt decomposition (SD)  
$\ket{\psi}_{ab} =  \sum_{n} \sqrt{\sigma_n}\ket{u_n v_n}_{ab}\to  \sum_{n} \sqrt{\sigma_n}\ket{nn}_{ab}$ 
(writing $\ket{u_n v_n}_{ab}\to\ket{nn}_{ab}$ for notational simplicity)  obtained by an SVD
\cite{NC:2000} of the amplitudes 
$c_{nm}\to C=U\,\Sigma\,V^\dag$, with $U$ and $V$ unitary, and  
$\Sigma=\trm{diagonal}\{\sqrt{\sigma_1},\sqrt{\sigma_2},\ldots\}>0$ and 
$\sum_n \sigma_n = 1=\sum_{n,m} |c_{nm}|^2 = \Tr[C\,C^\dag]$. 
We are interested in computing the Log Negativity (LN) given by $LN=\log_2(1+2\,\N)$ where the negativity 
$\N$ is the sum of the absolute values 
of the negative eigenvalues of the partial transpose  $\rho^\G_{ab}$ (on subsystem $b$)
of the pure state density matrix $\rho_{ab} = \ket{\psi}_{ab}\langle \psi|$.
Here, $M$ is the largest Fock state employed in each subsystem, set large enough so that $\{|c_{nm}|^2\le \epsilon\}$ is negligible for $n,m> M$. This allows us to  model both discrete and continuous variable states numerically.

\subsection{LN for diagonal wavefunctions or SD}\label{subsec:Neg:I}
As shown in Agarwal (\cite{Agarwal:2013}, p57) we can diagonalize  $\rho^\G_{ab}$ analytically as follows. First, let us consider a more general a pure state with diagonal complex amplitudes amplitudes $c_{nm} = c_n\,\delta_{m,n}$. 
We then have 
\bsub
\bea{Agarwal:p57:calc}
\hspace{-0.7in}
\rho_{ab} &=&  \sum_{n}\sum_{m}  c_{n}\, c^*_{m}\,\ket{nn}_{ab}\bra{mm}, \label{Agarwal:p57:calc:line:1} \\
\hspace{-0.7in}
\Rightarrow \rho^\G_{ab} &=&\sum_{n}\sum_{m}  c_{n}\, c^*_{m}\,\ket{nm}_{ab}\bra{mn}, \label{Agarwal:p57:calc:line:2} \\
\hspace{-0.7in}
&=& \sum_{n} | c_{n}|^2\,\ket{nn}_{ab}\bra{nn}  \no
\hspace{-0.7in}
&+& \half\sum_{n}\sum_{m\ne n}|c_{n}\, c_{m}| \left(e^{i\,\theta_{nm}}\ket{nm}_{ab}\bra{mn}+e^{-i\,\theta_{nm}}\ket{mn}_{ab}\bra{nm}\right),  \label{Agarwal:p57:calc:line:3}\no
\hspace{-0.7in}
&{}& 
\eea
\esub
where we have defined $c_{n}\, c^*_{m} = |c_{n}\, c_{m}| e^{i\,\theta_{nm}}$.
Now, the last term in \Eq{Agarwal:p57:calc:line:3} can be written as
\bsub
\bea{enm:states}
e^{i\,\theta_{nm}}\ket{nm}_{ab}\bra{mn}&+&e^{-i\,\theta_{nm}}\ket{mn}_{ab}\bra{nm} \label{enm:states:line:1}\no 
&=& 
\ket{e^+_{n,m}}\bra{e^+_{n,m}} -  \ket{e^-_{n,m}}\bra{e^-_{n,m}}, \\
\ket{e^\pm_{n,m\ne n}} &\equiv & \frac{1}{\sqrt{2}}\,\left( \ket{nm}\pm e^{-i\,\theta_{nm}}\,\ket{mn} \right). \quad \label{enm:states:line:2}
\eea
\esub
Note that the set of eigenstates $\{\ket{e^\pm_{n,m\ne n}},  \ket{nn}\}$ form a complete orthonormal set 
which diagonalizes $\rho^\G_{ab}$
and therefore allows us to read off the negative eigenvalue 
$\lambda^-_{n,m\ne n} = -\tfrac{1}{2}\,|c_n\,c_m|$ for $m\ne n$ in \Eq{enm:states:line:1}. 
We then conclude that
\bsub
\bea{LogNeg:I}
\ket{\psi}_{ab} &=& \sum_n c_n\,\ket{nn},  \\ \label{LogNeg:I:line:1}
 \Rightarrow  \N^I &\defn& \frac{1}{2} \sum_n\,\sum_{m\ne n} \, |c_n\,c_m|,  \label{LogNeg:I:line:2} \\
 \Rightarrow  LN &=& \log_2\left[ 1+ \sum_n\,\sum_{m\ne n} \, |c_n\,c_m| \right], \label{LogNeg:I:line:3} \\
 &=& \log_2\,\left[\left( \sum_n |c_n| \right)^2\right], \label{LogNeg:I:line:4}
\eea
\esub
where in the last line we have used $1= \sum_n |c_n|^2$.

Note that for the case of a SD, we have $c_n\to \sqrt{\sigma_n}$ so that we have for an arbitrary bipartite pure state
\be{LN:SD}
LN^{(SD)} \equiv\log_2\left[ \left( \sum_n  \sqrt{\sigma_n} \right)^2\right],
\ee
in terms of its Schmidt coefficients $\sigma_n$. 
However, as mentioned above, to obtain the Schmidt coefficients, one has to perform a SVD (or diagonalizaiton) on the pure state.

\subsection{LN for NmmN states}\label{subsec:Neg:I}
Another class of relevant states for our discussion are generalizations of the N00N states, which we denote as $NmmN$ states, of the form
$\ket{NmmN}\equiv \tfrac{1}{\sqrt{2}} \left(\ket{nm} + \ket{mn} \right)$.
Then it is straightforward to show that the PT of this state is given by
\bsub
\bea{Neg:NmmN}
\rho^\G &=& \frac{1}{2}
\Big(
\ket{nm}\bra{nm} + \ket{mn}\bra{nm} +
\ket{nn}\bra{mm} + \ket{mm}\bra{nn}
\Big),  \label{Neg:NmmN:line:1} \\
%
&\equiv&
\Big(
\ket{nm}\bra{nm} + \ket{mn}\bra{nm} +
\ket{f^+_{nm}}\bra{f^+_{nm}} - \ket{f^-_{nm}}\bra{f^-_{nm}} 
\Big), \qquad \label{Neg:NmmN:line:2} \\
%
 &{}& \ket{f^\pm_{n,m\ne n}} \equiv \frac{1}{\sqrt{2}} \left(\ket{nn}\pm \ket{mm} \right), \label{Neg:NmmN:line:3}
\eea
\esub
where we have defined the orthonormal set of eigenstates $\ket{f^\pm_{n,m\ne n}}$ in \Eq{Neg:NmmN:line:3}.
Note that the first two terms in \Eq{Neg:NmmN:line:1} are diagonal and orthogonal to  $\ket{f^\pm_{n,m\ne n}}$.
From \Eq{Neg:NmmN:line:2} we read off that $\lambda^-_{(NmmN)} = -1/2$ \tit{for all} $n,m$ and so
$\N = 1/2 \Rightarrow LN = \log_2(1+2\N) = 1$.

From the above, we note that if we have a pure state of the form
$\ket{\psi} = \sum_{n,m} c_{nm} \frac{1}{\sqrt{2}}  \left( \ket{n,m} + \ket{m,n} \right)$, for arbitrary complex $c_{nm}$,
i.e. a superposition of $NmmN$ states,
then each component state $\ket{NmmN}$ contributes a negativity of $1/2$ for each $n, m\ne n$ so that the total negativity is just
\bsub
\bea{Neg:II}
\ket{\psi} &=& \sum_{n,m} c_{nm} \frac{1}{\sqrt{2}}  \left( \ket{nm} + \ket{mn} \right), \\
%
\Rightarrow \N^{II} &\defn& \frac{1}{2}\sum_n \sum_{m\ne n} |c_{nm}\,c_{mn}|, \\
%
\Rightarrow LN^{II} &=& \log_2\left( 1+ 2\N^{II}\right) = \log_2(1+\sum_n \sum_{m\ne n} |c_{nm}\,c_{mn}|). \qquad
\eea
\esub

\section{Ansatz for an Entanglement Witness}\label{sec:Ansatz:Witness}
From the previous section we identify two types of states in the PT $\rho^\G_{ab}$ and their contributions to the negativity:
(without loss of generality and for notational simplicity, we treat $c_{nm}$ as real for now, 
and re-insert the absolute values at the very end of the discussion)

\bsub
\bea{Neg:I:and:Neg:II}
 \ket{nm}\bra{mn} +  \ket{mn}\bra{nm} &\leftrightarrow& \N^{I} \equiv \half \sum_{n,m\ne n} c_{nn}\,c_{mm} \label{Neg:I:and:Neg:II:line:1} \\
 \ket{nn}\bra{mm}  + \ket{mm}\bra{nn} &\leftrightarrow& \N^{II} \equiv \half\sum_{n,m\ne n} c_{nm}\,c_{mn}, \qquad\;\; \label{Neg:I:and:Neg:II:line:2}
\eea
\esub
where \Eq{Neg:I:and:Neg:II:line:1} comes from states eigenstates $\{\ket{e^\pm_{n,m\ne n}}\}$ from \Eq{enm:states:line:2}
leading to the negativity in \Eq{LogNeg:I:line:2} (where $c_n\to c_{nn}$), and 
\Eq{Neg:I:and:Neg:II:line:2} comes from eigenstates $\{\ket{f^\pm_{n,m\ne n}}\}$ from \Eq{Neg:NmmN:line:3} leading to the negativity in 
\Eq{Neg:I:and:Neg:II:line:2}. 

The main premise of our proposed entanglement witness (EW) is that these two types of terms 
constitute the majority contributions to the negativity for the general pure state $\ket{\psi}_{ab} = \sum_{n,m} c_{nm}$.
Below we argue that the LN can be lower bounded by the following expression built solely from the wavefunction amplitudes $c_{nm}$,
\bsub
\bea{LogNeg:Approx}
\N_{a}(\ket{\psi}_{ab})\defn \half\sum_n \sum_{m\ne n} \left| \trm{Det}\left(\begin{array}{cc} c_{nn} & c_{nm} \\c_{mn} &c_{mm}\end{array}\right)\right|, \label{LogNeg:Approx:line:1} \\
LN_{a}(\ket{\psi}_{ab}) = \log_2\left( 1+2\,\N_{a}\right). \label{LogNeg:Approx:line:2}
\eea
\esub
At first glance \Eq{LogNeg:Approx:line:1} looks a bit odd, since if one were to take $\N_{a}=\N^I+\N^{II}$ 
(again, subscript $a$ for \tit{approximate}) one would expect to sum over the  permanent $\mathcal{P} \equiv | c_{nn}\,c_{mm} + c_{nm}\,c_{mn}|$ as opposed to the determinant 
$\mathcal{D}=| c_{nn}\,c_{mm} - c_{nm}\,c_{mn}|$ \cite{meyer:wallach:comment,Meyer_Wallach:2002}. 
However, using $\mathcal{D}$ enforces $\N_{a}=0$ on separable (product) pure states (i.e. the Det is identically zero if $c_{nm} = a_n\,b_m$). 
As mentioned previously, the set of eigenstates $\{\ket{e^\pm_{n,m\ne n}}, \ket{nn}\}$ giving rise to  $\N^I$
form an orthonormal subset, while the set of eigenstates $\{\ket{f^\pm_{n,m\ne n}}, \ket{nn}\}$ giving rise to  $\N^{II}$ do not. 
Thus, $\N^{II}$ is being ``overcounted" in $\N_{a}=\N^I+\N^{II}$ and we have found that 
$\N_{a}\defn\N^I-\N^{II}$ produces much better results. 
Additionally, the use of the latter minus sign assures that if $c_{nm} =  \tfrac{1}{\sqrt{d}}\,\delta_{nm}$ 
then we obtain $\ket{\psi}_{ab} = \tfrac{1}{\sqrt{d}}\sum_{n}\ket{nn}_{ab} \equiv \ket{Bell^{(d)}}_{ab}$ the 
$d$-dimensional maximally entangled Bell state, for which $LN_{a}=LN_{e}=\log_2 d$.

\subsection{$\boldsymbol{LN_{a}}$: Results}\label{subsec:results}
\begin{figure}[h]
\includegraphics[width=3.0in,height=2.25in]{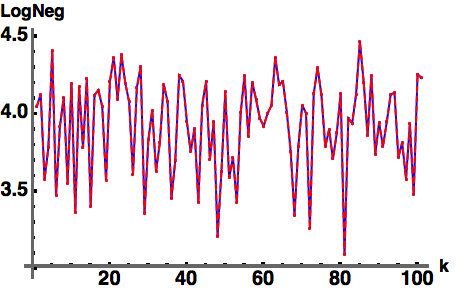} 
\caption{$LN(\ket{\psi}_{ab})$: (blue) exact numerical, (red) approximate from \Eq{LogNeg:Approx:line:2}, for pure state with 
$\ket{\psi}_{ab} = \sum_{n,m=0}^{M}\,c_{nm} \ket{nm}_{ab} $ 
with $c_{nm}$ generated as a positive $(M+1)\times(M+1)$ Hermitian matrix flattened (stripped by rows) into an 
$(M~+~1)^2$ column vector, with $c_{nm} = \left(U\,\rho_{diag}\,U^\dag\right)_{nm}$. Here, $\rho_{diag}$ is a random positive diagonal matrix such that $\sum_{n=0}^M (\rho_{diag})_{nn} = 1$, and $U$ is a uniformly generated (via the Haar measure, see \App{app:Haar:purity}) $(M+1)\times(M+1)$ random unitary matrix. 
 (Note: $\rho_{diag}$ is chosen as the absolute value \tit{squared} of a random row of another randomly chosen unitary $U'$).
In this figure,  $M=40$, with ($k$) $100$ samples (results are the same regardless the number of samples chosen).
}\label{fig:cnm:random:M40_U2}
\end{figure}

Surprisingly, we have found that \Eq{LogNeg:Approx:line:1} and \Eq{LogNeg:Approx:line:2} produce \tit{exactly} $LN_{e}$ (obtained from numerically computed eigenvalues) of pure states for which 
$c_{nm}$ is a \tit{positive Hermitian matrix} with both deterministic,  or random entries, as shown in 
\Fig{fig:cnm:random:M40_U2}. 
Further, for the case when $c_{nm}$ is a positive Hermitian matrix $C\to\rho>0$ 
we can derive (see \Eq{tildeN:A:Hermitian:TrAneq1:line1} and \Eq{tildeN:A:Hermitian:TrAneq1:line2})
the \tit{non-trivial} result 
\bsub
\bea{LN:Approx:cnm:random:U2}
\hspace{-0.6in}
C\to\rho>0 
\Rightarrow \N_{a}&=& \half\,\left(\frac{\left(\Tr[\rho]\right)^2}{\Tr[\rho^2]}-1\right)=\N_{e}, \qquad \label{LN:Approx:cnm:random:U2:line:1} \\
\hspace{-0.6in}
\Rightarrow LN_{a} &=& \log_2\left(\frac{\left(\Tr[\rho]\right)^2}{\Tr[\rho^2]}\right)= LN_{e}
\overset{\Tr[\rho]=1}{\longrightarrow}-\log_2(\Tr[\rho^2]), \qquad\;\, 
 \label{LN:Approx:cnm:random:U2:line:2} 
\eea
\esub
where the second equalities in the above \tit{cannot} be derived analytically, but rather 
are demonstrated numerically from plots such as \Fig{fig:cnm:random:M40_U2}.
Here, $C\to\rho=\left(U\,\rho_{diag}\,U^\dag\right) >0$ which is then flattened (stripped by rows) into an $(M+1)^2\times 1$ column vector 
representing $\ket{\psi}_{ab}$.
$\rho_{diag}$ is chosen as a random positive diagonal matrix such that $\sum_{n=0}^M (\rho_{diag})_n~=~1$, and $U$ is a uniformly generated (via the Haar measure) $(M+1)\times(M+1)$ random unitary matrix. 
 (Note: $\rho_{diag}$ is chosen as the absolute value \tit{squared} of a random row of another randomly chosen unitary $U'$).
 
\begin{figure}[h]
\includegraphics[width=3.0in,height=2.25in]{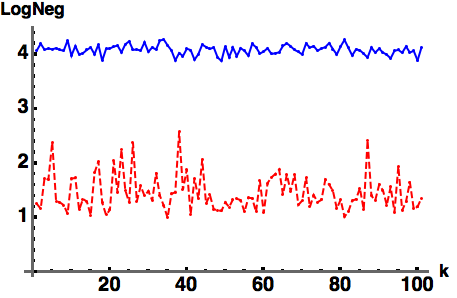} 
\caption{$LN(\ket{\psi}_{ab})$: (blue) exact numerical , (red) approximate from \Eq{LogNeg:Approx:line:2}. Same as 
\Fig{fig:cnm:random:M40_U2} except now $\rho_{diag}\to C_{diag}$ is chosen simply as a random row of separately generated randomly unitary $U'$ (vs the absolute value squared of a random row of $U'$) so that it has random complex entries. 
Here, we take $M=40$ with $100$ samples.
For any number of samples we always find that $LN_{a}(\ket{\psi}_{ab})\le LN_{e}(\ket{\psi}_{ab})$, i.e. 
$LN_{a}(\ket{\psi}_{ab})$ acts as proper lower bound to $LN_{e}(\ket{\psi}_{ab})$.
}\label{fig:cnm:random:M40_U1}
\end{figure}

 In \Fig{fig:cnm:random:M40_U1} we relax the positivity condition on $C$ and change $\rho_{diag}\to C_{diag}$ to be chosen simply as a random row of separately generated randomly unitary $U'$ (vs the absolute value squared of a random row of $U'$ for the previous case of $C\to\rho>0$) so that $C_{diag}$ has random complex entries.
 For any number of samples we always find that $LN_{a}(\ket{\psi}_{ab})\le LN_{e}(\ket{\psi}_{ab})$, i.e. 
$LN_{a}(\ket{\psi}_{ab})$ acts as proper lower bound to $LN_{e}(\ket{\psi}_{ab})$. 
This behavior in \Fig{fig:cnm:random:M40_U1} is the same if we simply take $C=\{c_{nm}\}$ as a matrix of random complex  entries.
This represents one of the main results of this work. In the following subsection we provide a plausibility argument for \Eq{LogNeg:Approx:line:2} indicating the logic of our ``derivation."

\begin{figure}[ht]
\begin{tabular}{c}
\includegraphics[width=2.5in,height=1.5in]{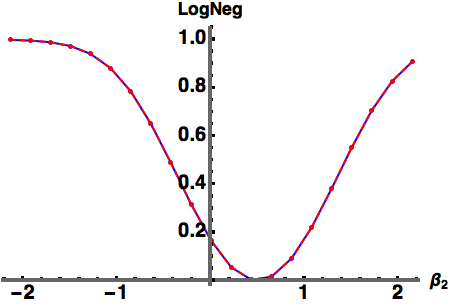} \\
\includegraphics[width=2.5in,height=1.5in]{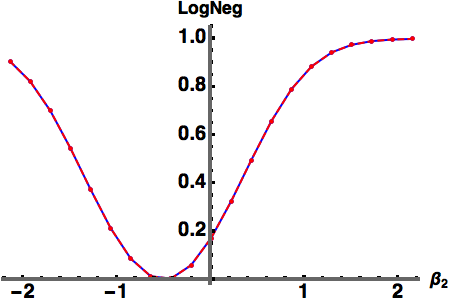}
\end{tabular}
\caption{$LN(\ket{\Psi}_{ab})$: (blue) exact numerical, (red) approximate,
for the state
$\ket{\Psi}_{ab}=\textsf{N}\left( \ket{\beta_1}_a\ket{\beta_2}_b +  \ket{\beta_2}_a\ket{\beta_1}_b\right)/\sqrt{2}$.
(top) $\beta_1=0.5, M=20$,
(bottom) $\beta_1=-0.5, M=20$.
The state is separable when $\beta_2=\beta_1$ where $LN\to0$.
}\label{fig:beta1221:M20}
\end{figure}

As discussed previously, the use of the Det in \Eq{LogNeg:Approx:line:1} enforces  $\N_{a}=0$ on separable (product) pure states since the Det is identically zero if $c_{nm} = a_n\,b_m$. This is nicely illustrated for the case of pure states of the form
$\ket{\Psi}_{ab} = \textsf{N}\left( \ket{\psi}_a\ket{\phi}_b +  \ket{\phi}_a\ket{\psi}_b\right)/\sqrt{2}$, where
$ \textsf{N}~=~\left(1+|\bra{\psi}\phi\rangle|^2 \right)^{-1/2}$. 
In \Fig{fig:beta1221:M20} we show the superposition of two coherent states
$\ket{\Psi}_{ab} = \textsf{N}\left( \ket{\beta_1}_a\ket{\beta_2}_b +  \ket{\beta_2}_a\ket{\beta_1}_b\right)/\sqrt{2}$.
The state is separable when $\beta_2=\beta_1$ where $LN\to0$ (both exact and approximate).
\begin{figure}[ht]
\begin{tabular}{cc}
\hspace{-0.2in}
\includegraphics[width=2.05in,height=1.5in]{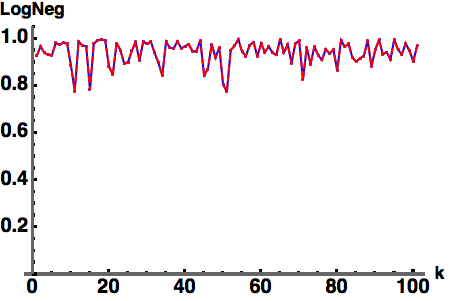} &
\hspace{-0.1in}
\includegraphics[width=2.05in,height=1.5in]{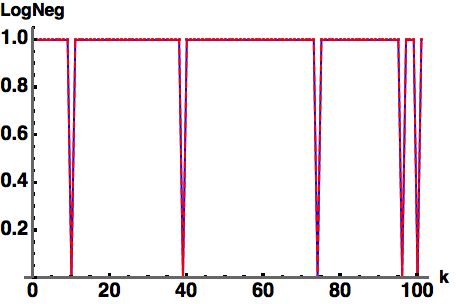}
\end{tabular}
\caption{$LN(\ket{\Psi}_{ab})$: (blue) exact numerical, (red) approximate,
for the state
$\ket{\Psi}_{ab}=\textsf{N}\left( \ket{\psi}_a\ket{\phi}_b +  \ket{\phi}_a\ket{\psi}_b\right)/\sqrt{2}$ for $M=20$ with
$\ket{\psi}$ and $\ket{\phi}$ taken as two random rows of 
(left) two different random unitary matrices,
(right) the same unitary matrix
The state is separable when $\ket{\psi}$ and $\ket{\phi}$ are the same random vector, where $LN\to0$.
}\label{fig:psiphiphipsi:M20:different:same:unitary}
\end{figure}

Another way to show that \Eq{LogNeg:Approx:line:2} captures separable pure state is shown in 
\Fig{fig:psiphiphipsi:M20:different:same:unitary}
In the left figure $\ket{\psi}_a$ and $\ket{\phi}_b$ taken as two random rows of two different random unitary matrices, 
while in the right plot, $\ket{\psi}_a$ and $\ket{\phi}_b$ taken as two random rows of the same random unitary matrix. In the latter, when the same row is chosen for both  $\ket{\psi}_a$ and $\ket{\phi}_b$, the state is separable and $LN\to0$.

\subsection{Plausibility argument for \Eq{LogNeg:Approx:line:1} and \Eq{LogNeg:Approx:line:2}}\label{subsec:derivation}
In this section we present a plausibility argument that led to the ``derivation`` of the Negativity in \Eq{LogNeg:Approx:line:1} and hence the LogNegativity in \Eq{LogNeg:Approx:line:2}.

We begin by considering a general pure state $\ket{\psi}_{ab} = \sum_{n,m} c_{nm}\ket{n,m}_{ab}$.
The pure state density matrix is given by  
$\rho_{ab}=\ket{\psi}_{ab} \bra{\psi}= \sum_{n,m} \sum_{n',m'} c_{nm}\, c^*_{n'm'}\ket{n,m}_{ab}\bra{n'm'}$.
Without loss of generality, and for ease of notation, we treat $c_{nm}$ as real in this subsection, and put in appropriate absolute values at the end of the calculation. Also, we drop the $ab$ subscript for now.
We now write $\rho_{ab}$ as
\bwt
\bsub
\bea{rhoab}
\rho_{ab} &=& \sum_{n,n'} c_{nn}\,c_{n'n'} \ket{nn}\bra{n'n'} + 
\frac{1}{4}  \sum_{n,m\ne n} \, \sum_{n',m'\ne n'}  c_{nm}\,c_{n'm'}\,\left(\ket{nm}+\ket{mn} \right)\left(\bra{n'm'} + \bra{m'n'} \right), \\
&+& \frac{1}{2}  \sum_{n} \sum_{n',m'\ne n'}  c_{nn}\,c_{n'm'}\,\ket{nn}\left(\bra{nm}+\bra{mn} \right)
+   \frac{1}{2}     \sum_{n,m\ne n} \sum_{n'} c_{nm}\,c_{n'n'}\,\left(\ket{nm}+\ket{mn} \right)\bra{nn}.
\eea
\esub
The PT $\rho^\G_{ab}$ is then given by 
\bsub
\bea{rhoab:PT}
\hspace{-0.5in}
\rho^\G_{ab} &=& \sum_{n,m} c_{nn}\,c_{mm} \ket{nm}\bra{mn} + 
\frac{1}{4}  \sum_{n,m\ne n} \, \sum_{n',m'\ne n'}  c_{nm}\,c_{n'm'}\,
\left(
\ket{nm'}\bra{n'm} + \ket{mn'}\bra{m'n} +\ket{nn'}\bra{m'm} +\ket{mm'}\bra{n'n} 
\right), \label{rhoab:PT:line:1} \\
\hspace{-0.5in}
&+& \frac{1}{2}  \sum_{n} \sum_{n',m'\ne n'}  c_{nn}\,c_{n'm'}\,\left(\ket{nm'}\bra{n'n}+\ket{nn'}\bra{m'n} \right)
+   \frac{1}{2}     \sum_{n,m\ne n} \sum_{n'} c_{nm}\,c_{n'n'}\,\left(\ket{nn'}\bra{n'm}+\ket{mn'}\bra{n'n} \right). \label{rhoab:PT:line:2}
\eea
\esub
\ewt
The first summation in \Eq{rhoab:PT:line:1} gives rise to the Negativity $\N^{I} = \half \sum_{n,m\ne n} |c_{nn}\,c_{mm}|$
(putting back in the absolute values)
as discussed in \Eq{Neg:I:and:Neg:II:line:1} using the eigenstates $\ket{e^\pm_{n,m\ne n}}$.

The ansatz we employ is that the Negativity is dominated by terms of the form (I): $\ket{nm}\bra{mn} + \ket{mn}\bra{nm}$, 
giving rise to eigenstates $\ket{e^\pm_{n,m\ne n}}$,
and terms 
(II): $\ket{nn}\bra{mm} + \ket{mm}\bra{nn}$,
giving rise to eigenstates $\ket{f^\pm_{n,m\ne n}}$,
in the PT $\rho^\G_{ab}$. We therefore approximate the Negativity by only considering ``matching terms" 
of type $I$ and $II$ in $\rho^\G_{ab}$.

Thus, in the second double summation in \Eq{rhoab:PT:line:1} we only consider the terms (i) $n'=n, m'=m$
which give rise to terms of the form
$\tfrac{1}{4}\sum_{n,m\ne n}c_{nm}\,c_{nm}\,
\big(
\ket{nm}\bra{nm} + \ket{mn}\bra{mn} 
$
$
+\ket{nn}\bra{mm} +\ket{mm}\bra{nn} 
\big)$.
The first two terms of this expression are diagonal, while the last term terms gives rise to the Negativity 
$\half \N^{II} = \tfrac{1}{4} \sum_{n,m\ne n} |c_{nm}\,c_{mn}|$
as discussed in \Eq{Neg:I:and:Neg:II:line:2} using the eigenstates $\ket{f^\pm_{n,m\ne n}}$.

However, in the same term in the previous paragraph, we could also consider the case (ii) $n'=m, m'=n$ leading to
terms of the form 
$\tfrac{1}{4}\sum_{n,m\ne n}c_{nm}\,c_{nm}\,
\big(
\ket{nn}\bra{mm} + \ket{mm}\bra{nn} 
$
$
+\ket{nm}\bra{nm} +\ket{mn}\bra{mn} 
\big)$.
The last two terms of this expression are diagonal, while the first two again contribute a Negativity of 
$\half \N^{II} = \tfrac{1}{4} \sum_{n,m\ne n} |c_{nm}\,c_{mn}|$, as in the previous paragraph.
As part of our ansatz, 
we conjecture that the terms in \Eq{rhoab:PT:line:2} contribute no (significant) terms to the Negativity.

Thus, the Negativities that we have so far are $\N \approx \N^{I} + \N^{II}$ which we can write as
\bsub
\bea{NegI:NegII:derivation}
\N &\approx& \N^{I} + \N^{II}, \label{NegI:NegII:derivation:line:1} \\
%
 &=&\half\sum_{n,m\ne n} |c_{nn}\,c_{mm}| + \half\sum_{n,m\ne n} |c_{nm}\,c_{mn}|, \label{NegI:NegII:derivation:line:2} \\
 %
%
&\ge& 
\half\sum_n \sum_{m} \left| \trm{Det}\left(\begin{array}{cc} c_{nn} & c_{nm} \\c_{mn} &c_{mm}\end{array}\right)\right|
\defn \N_{a}, \label{NegI:NegII:derivation:line:4} \qquad
\eea
\esub
where in the last line we have simply used the inequality that $ |a| + |b| \ge |a-b|$ with $a\to c_{nn}\,c_{mm}$ and $b\to c_{nm}\,c_{mn}$.
As discussed before, using $\N \approx \N^{I} + \N^{II}$ in \Eq{NegI:NegII:derivation:line:1} overestimates the Negativities because the set of eigenstates $\{ \ket{f^\pm_{n,m\ne n}}, \ket{nn}\}$ does not form an orthogonal subset. 
Thus, we instead use the lowerbound $\N\approx\N_{a}$ given by the determinant expression in \Eq{NegI:NegII:derivation:line:4}. $\N_{a}$  is also the same expression as in \Eq{LogNeg:Approx:line:1}, giving rise to the $LN$ given in  \Eq{LogNeg:Approx:line:2}.

\begin{figure}[!ht]
\begin{tabular}{cc}
\hspace{-0.2in}
\includegraphics[width=2.15in,height=1.5in]{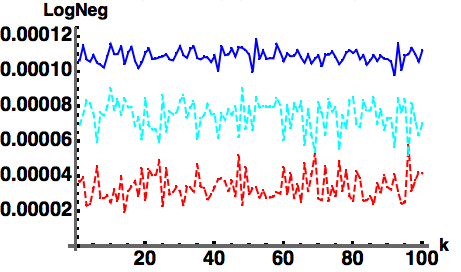} &
\hspace{-0.2in}
\includegraphics[width=2.15in,height=1.5in]{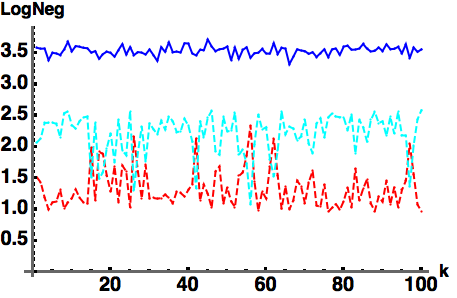}  \\
\hspace{-0.2in}
\includegraphics[width=2.15in,height=1.5in]{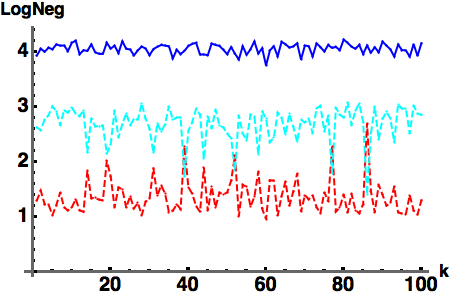} &
\hspace{-0.2in}
\includegraphics[width=2.15in,height=1.5in]{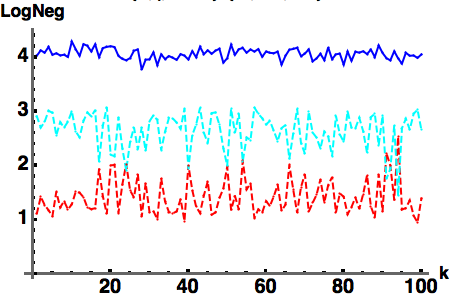}  
\end{tabular}
\caption{
(blue) $LN_{e}$, 
(red) $LN_{a}$ and 
(cyan) $LN_{e}-LN_{a}$ for 
$\ket{\psi}_{ab}=\sum_{nm} c_{nm} \ket{nm}_{ab}$ with
$\{c_{nm}\}\equiv\rho(\eta)=\ket{\phi_0}\bra{\phi_0} + \eta\,\sigma$ for $M=20$. 
Here,  $\ket{\phi_0}=\sum_n (\phi_0)_n\,\ket{n}$ is an arbitrary $d$ dimensional complex vector, such that  
$LN_{a}(\rho(0))=LN_{e}(\rho(0))=0$ and $\sigma$ is a random complex matrix, and $\eta$ an arbitrary real scale factor.
(top left) $\eta=10^{-5}$,
(top right) $\eta=1$,
(bottom left) $\eta=10$
(bottom right) $\eta=100$.
}\label{fig:rho:eq:rho0:plus:eta:sigma:M20}
\end{figure}
\begin{figure}[!ht]
\begin{tabular}{cc}
\hspace{-0.2in}
\includegraphics[width=2.15in,height=1.5in]{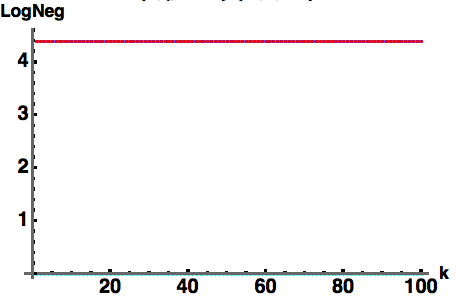} &
\hspace{-0.2in}
\includegraphics[width=2.15in,height=1.5in]{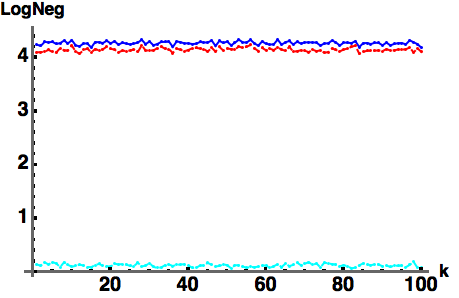}  \\
\hspace{-0.2in}
\includegraphics[width=2.15in,height=1.5in]{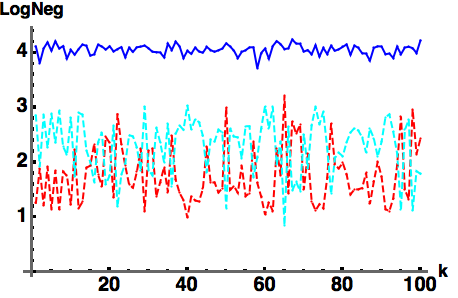} &
\hspace{-0.2in}
\includegraphics[width=2.15in,height=1.5in]{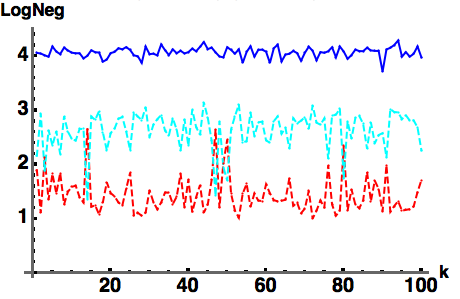}  
\end{tabular}
\caption{
(blue) $LN_{e}$, 
(red) $LN_{a}$ and 
(cyan) $LN_{e}-LN_{a}$ for 
$\ket{\psi}_{ab}=\sum_{nm} c_{nm} \ket{nm}_{ab}$ with
$\{c_{nm}\}\equiv\rho(\eta)=\mathbb{I}/d + \eta\,\sigma$ for $M=20$ 
where  $\ket{\phi_0}=\sum_n c_n\,\ket{n}$ is an arbitrary $d$ dimensional complex vector, such that  
$LN_{a}(\rho(0))=LN_{e}(\rho(0))=1$ and $\sigma$ is a random complex matrix, and $\eta$ an arbitrary real scale factor.
(top left) $\eta=0$,
(top right) $\eta=0.1$,
(bottom left) $\eta=1$
(bottom right) $\eta=10$.
}\label{fig:rho:eq:MMS:plus:eta:sigma:M20}
\end{figure}

While the above does not constitute a formal proof or a proper derivation $\N_{a}$, it is a plausibility argument borne out by numerical evidence. $\N_{a}$ and $LN_{a}$ in  \Eq{LogNeg:Approx:line:1} and  \Eq{LogNeg:Approx:line:2} also argue that the Negativity in a general pure state is dominated by the states $\ket{nn}_{ab}\bra{mm} + \ket{mm}_{ab}\bra{nn}$ and 
 $\ket{nm}_{ab}\bra{mn} + \ket{mn}\bra{nm}$ found within $\ket{\psi}_{ab} = \sum_{nm} c_{nm}\,\ket{nm}_{ab}$, giving rise in the PT $\rho^\G_{ab}$ to the eigenstates and Negativities $\{\ket{e^\pm_{n,m\ne n}}_{ab}, \ket{f^\pm_{n,m\ne n}}_{ab} \}$ and $\{\N^{I}, \N^{II}\}$, respectively. 

Finally, we return to the case when 
 $c_{nm} =  c_n\,c^*_m$ is separable so that $LN_{a}=LN_{e}=0$.
Let us now consider $C=\{c_{nm}\} \equiv \rho/\sqrt{Tr[\rho^2]}$ where we define
$\rho=\rho_0/\sqrt{Tr[\rho_0^2]} + \eta\,\sigma$, with $\rho_0$ and $\sigma$ Hermitian.
If we further define $\rho(\eta=0)=\rho_0 \equiv \ket{\phi_0}\bra{\phi_0}$ where 
$\ket{\phi_0}=\sum_n c_n\,\ket{n}$ is an arbitrary $d$ dimensional complex vector, (such that  $LN_{a}(\rho_0)=LN_{e}(\rho_0)=0$), then $\rho(\eta)$ represents and increasing random deviation away from the separable case.
In \Fig{fig:rho:eq:rho0:plus:eta:sigma:M20} we show plots of 
$LN_{e}$, $LN_{a}$ and $LN_{e}-LN_{a}$
for $\rho(\eta)$,
such that $LN_{a}(\rho(0))=LN_{e}(\rho(0))=0$, 
with $M=20$ for increasing values of $\eta$.
We see that $LN_{a}$ acts as a lower bound for $LN_{e}$.
In \Fig{fig:rho:eq:MMS:plus:eta:sigma:M20} we show plots of 
$LN_{e}$, $LN_{a}$ and $LN_{e}-LN_{a}$
for $\rho(\eta)=\mathbb{I}/d + \eta\,\sigma$,
such that $LN_{a}(\rho(0))=LN_{e}(\rho(0))=1$,
with $M=20$ for increasing values of $\eta$.
We again see that $LN_{a}$ acts as a lower bound for $LN_{e}$.
Similar behavior occurs for arbitrary values of~$M$.

 \subsection{Analytic derivation of $\boldsymbol{LN_{a}(\ket{\psi}_{ab})\equiv LN_{e}(\ket{\psi}_{ab})}$ for the bipartite case of two  qubits, $\boldsymbol{d=2}$}\label{subsec:2qubits}
 For the case of two qubits ($M=1, d=2$),  take $C=\{c_{nm}\}$ to be an arbitrary
 $2\times 2$ complex matrix, for which we can analytically obtain the eigenvalues of the 
 $2^2\times 2^2$ partial transpose 
 $\rho_{ab}^\G$,  \Eq{rhoab:PT:line:2}, in terms of the $c_{nm}$. 
 After using the normalization of the state $\Tr[C C^\dag]$
 $=|c_{00}|^2 + |c_{01}|^2 +|c_{10}|^2 +|c_{11}|^2=1$,   
 the eigenvalues $\{\lambda_1, \lambda_2,\lambda_+,\lambda_-\}$
 of  $\rho_{ab}^\G$ can be written as
 \bsub
 \bea{evals:2qubits}
 \lambda_1 &=& |c_{00}\,c_{11} - c_{01}\,c_{10}| = -\lambda_2, \label{evals:2qubits:line1} \label{evals:2qubits:line1} \\ 
 %
 \lambda_\pm &=& \half\,\big(1 \pm \sqrt{1-4\,\lambda_1^2}\, \big).\qquad \label{evals:2qubits:line2}
 \eea
\esub
Thus, $\lambda_2\le 0$ and contributes to the Log Negativity.
Since $\rho^\G_{ab}$ is Hermitian with real eigenvalues, the radical under the square root must be positive
$1-4\,\lambda_1\ge 0$, requiring that $\lambda_1\le1/2$.
%
\begin{figure}[!ht]
\begin{tabular}{cc}
\hspace{-0.5in}
\includegraphics[width=2.0in,height=1.5in]{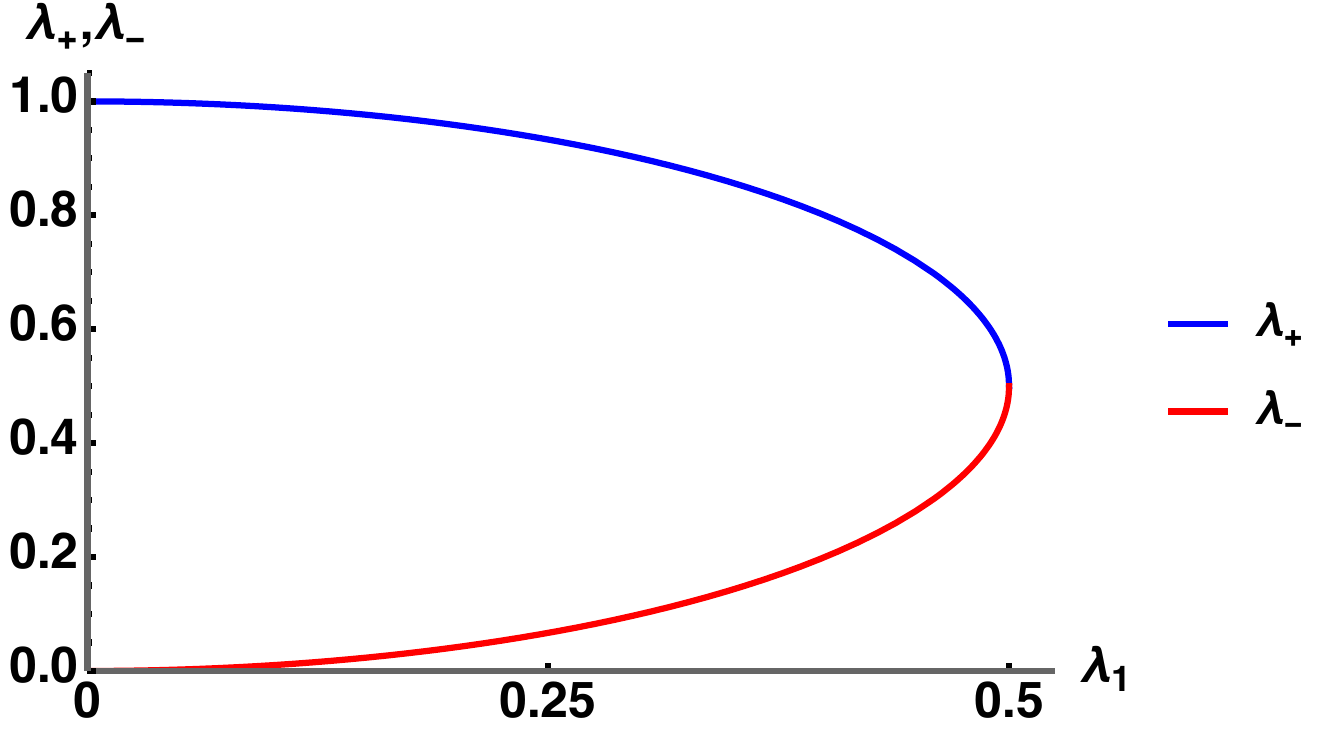} &
\includegraphics[width=1.85in,height=1.5in]{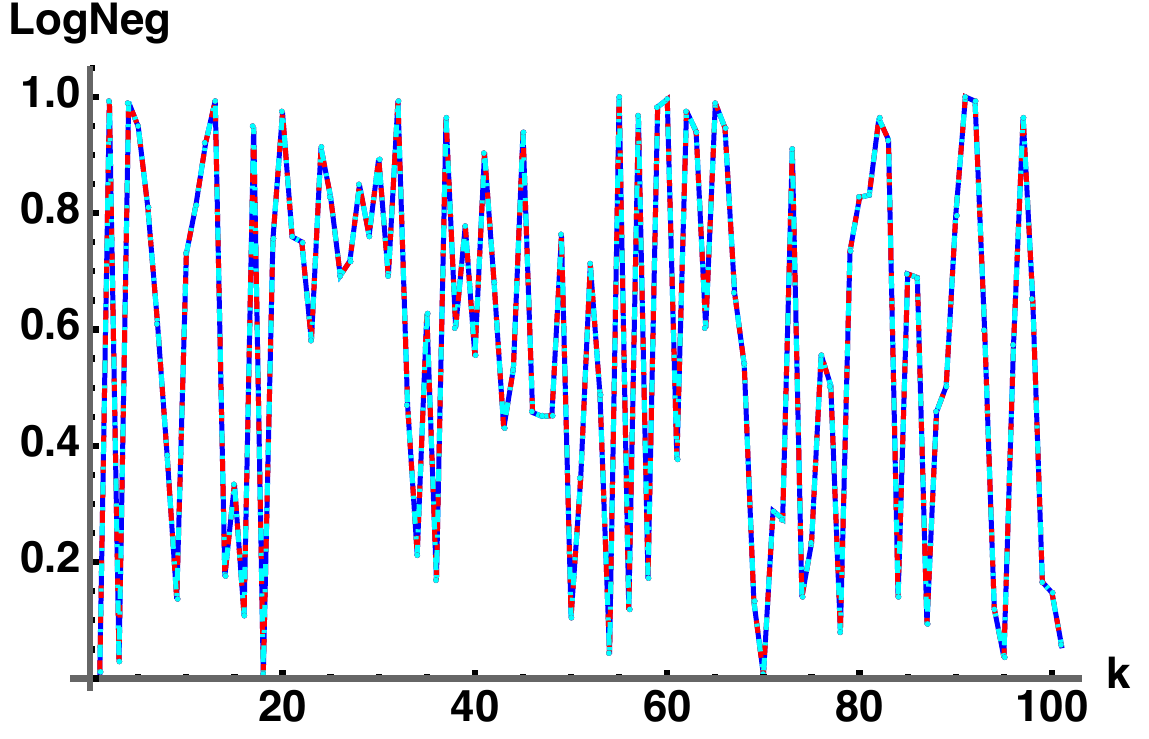} 
\end{tabular}
\caption{
(left)Plot of eigenvalues 
$
\lambda_\pm = 
\half\,
\big(
1 \pm \sqrt{1-4\,\lambda^2_1}\,
\big)
$ 
 from \Eq{evals:2qubits:line2}.
 (right) $LN_{e}$ (blue), $LN_{a}$ (red) from \Eq{LogNeg:Approx:line:2}, 
 and $LN$ (cyan) using analytic eigenvalue $\lambda_2$ in \Eq{evals:2qubits:line2} for 
 the case of two qubits, $d=2$.
}\label{fig:lambda:plus:minus:vs:lambda1}
\end{figure}
In \Fig{fig:lambda:plus:minus:vs:lambda1}(left) we plot $\lambda_\pm$. 
In \Fig{fig:lambda:plus:minus:vs:lambda1}(right) we plot 
$LN_{e}$ (blue), $LN_{a}$ (red) from \Eq{LogNeg:Approx:line:2}, 
 and $LN$ (cyan) using the analytic eigenvalue $\lambda_1=|\lambda_2|$ 
 in \Eq{evals:2qubits:line2}, for 
 the bipartite state of two qubits, $d=2$.
The end result is that for the   case of the pure bipartite case of two qubits there is only a single negative eigenvalue of the PT that contributes to the Negativity, and it has the precise form of $\N_a$ as given in \Eq{LogNeg:Approx:line:1}, which has two terms in the sum $(n,m) = \{(0,1),(1,0)\}$ summing to $|\lambda_2|=\lambda_1$.
For higher dimensions $d>2$ it is then somewhat unexpected and surprising that $\N_a$ in \Eq{LogNeg:Approx:line:1}  produces the exact Negativity $\N_e$ when $C=C^\dag$ of dimension $d$.

\subsection{Comparison with Linear Entropy}
The archetypal measure of  entanglement of bipartite pure states that does \tit{not} involve the computation of eigenvalues is the linear entropy $LE(\ket{\psi}_{ab})=1-\Tr[\rho^2_a]$ where $\rho_a = \Tr_b[\ket{\psi}_{ab}\bra{\psi}]$ is the reduced density matrix obtained by tracing out over one of the subsystems. $LE=0$ when $\ket{\psi}_{ab}$ is separable and 
$LE=1-\tfrac{1}{d}$ when 
$\rho_{ab}=\ket{\psi}_{ab}\bra{\psi}$ (a $d^2\times d^2$ matrix) is maximally entangled, and hence $\rho_{a}$ (a $d\times d$ matrix) is maximally mixed.
\begin{figure}[!ht]
\begin{tabular}{cc}
\hspace{-0.2in}
\includegraphics[width=2.15in,height=1.5in]{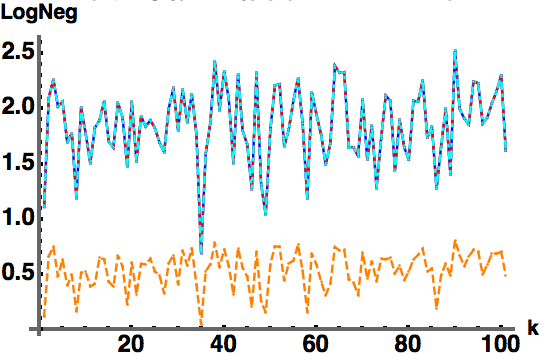} &
\hspace{-0.2in}
\includegraphics[width=2.15in,height=1.5in]{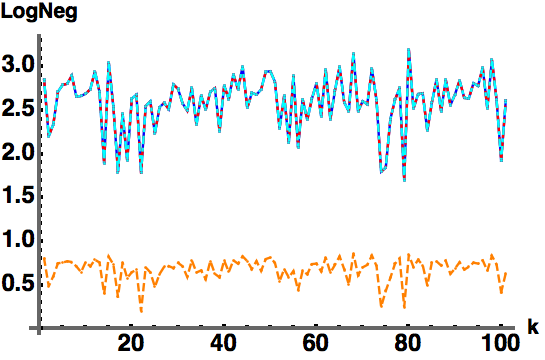}  \\
\hspace{-0.2in}
\includegraphics[width=2.15in,height=1.5in]{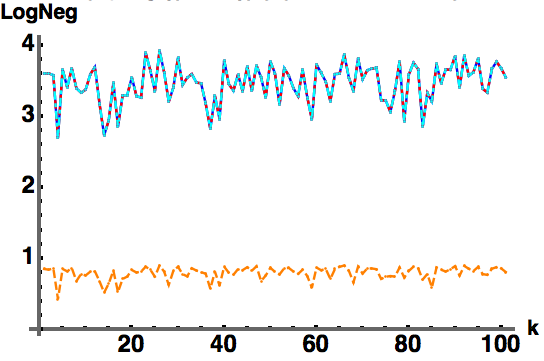} &
\hspace{-0.2in}
\includegraphics[width=2.15in,height=1.5in]{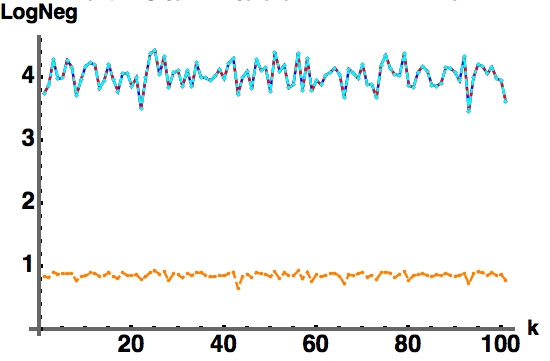}  
\end{tabular}
\caption{(blue) 
$LN_{e}$, 
(red) $LN_{a}$,
(cyan)~$LN_{a}^{variation}$, and 
(orange) $LE$ 
when
$\{c_{nm}\}\equiv\rho>0$ is a positive Hermitian $d\times d$ matrix (with $d=M+1$):
(top left) $M=5$,
(top right) $M=10$,
(bottom left) $M=20$
(bottom right) $M=30$.
}\label{fig:LNE:LNA:LNAvar:LE:Ceqrho:M5:10:20:30}
\end{figure}
In \Fig{fig:LNE:LNA:LNAvar:LE:Ceqrho:M5:10:20:30} we plot $LN_{e}$,  $LN_{a}$, 
and a variation on the approximate Negativity formula  \Eq{LogNeg:Approx:line:2} given by 
$LN_{a}^{variation} = \log_2\left(1+\big| \Tr[C] \Tr[C^\dag] - \Tr[C C^\dag]\big|\right)$ 
(see \Eq{tildeN:calc:C}), and 
the linear entropy $LE$, when $\{c_{nm}\}=C\to\rho>0$ is a positive Hermitian $d\times d$ matrix 
(with $d=M+1$) for $M=\{5,10,20,30\}$. In this case, both $LN_{a}$ and $LN_{a}^{variation}$ identically equal  $LN_{e}$. The $LE$  is plotted below (in orange) and acts as a lower bound to $LN_{e}$.

\begin{figure}[!ht]
\begin{tabular}{cc}
\hspace{-0.5in}
\includegraphics[width=2.15in,height=1.5in]{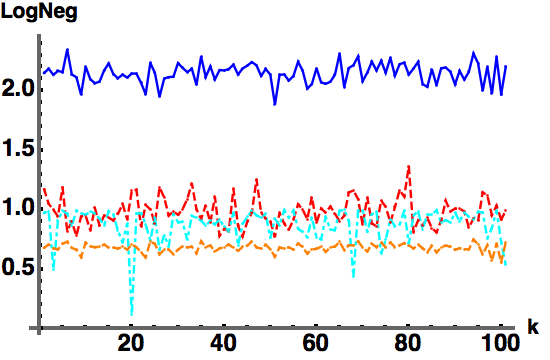} &
\hspace{-0.5in}
\includegraphics[width=2.15in,height=1.5in]{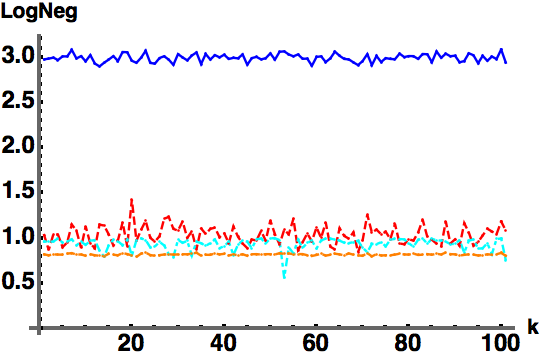}  \\
\hspace{-0.5in}
\includegraphics[width=2.15in,height=1.5in]{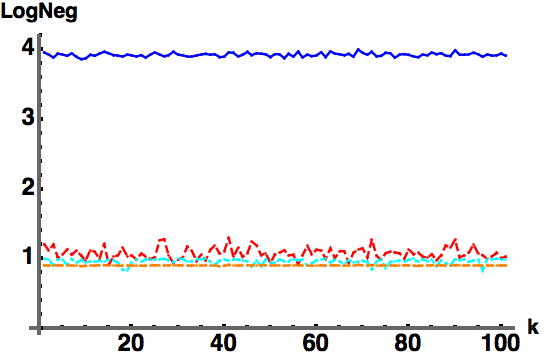} &
\hspace{-0.5in}
\includegraphics[width=2.15in,height=1.5in]{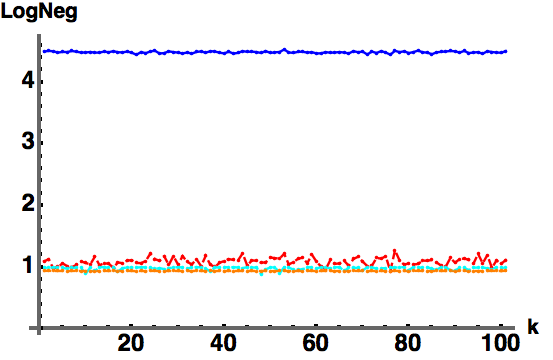}  
\end{tabular}
\caption{
(blue) $LN_{e}$, 
(red) $LN_{a}$,
(cyan)~$LN_{a}^{variation}$, and 
(orange) $LE$ 
when
$\{c_{nm}\}=C$ is a random complex $d\times d$ matrix (with $d=M+1$):
(top left) $M=5$,
(top right) $M=10$,
(bottom left) $M=20$
(bottom right) $M=30$.
(Compare with \Fig{fig:LNE:LNA:LNAvar:LE:Ceqrho:M5:10:20:30}).
}\label{fig:LNE:LNA:LNAvar:LE:CeqRandomComplexMatrix:M5:10:20:30}
\end{figure}
\Fig{fig:LNE:LNA:LNAvar:LE:CeqRandomComplexMatrix:M5:10:20:30} is identical to \Fig{fig:LNE:LNA:LNAvar:LE:Ceqrho:M5:10:20:30} except we now allow 
$\{c_{nm}\}=C$ to be a random complex $d\times d$ matrix.
We see that $LN_{a}$ acts as a better lower bound than $LN_{a}^{variation}$, which is in turn a better lower bound than the $LE$. 
As the dimension $d$ increases, the latter three measures are observed to converge to approximately the same lower bound values, with $LN_{a}$ still performing as a slightly better lower bound.

 \section{An attempt to extend $\boldsymbol{\N_{a}}$ to general density matrices}\label{sec:dms}
 Extending the Negativity and LogNegativity from  \Eq{LogNeg:Approx:line:1} and  \Eq{LogNeg:Approx:line:2} appropriate for pure states to mixed states is a non-trivial task.
 In principle the extension of an entanglement measure to mixed states is easy to state, but computationally involved to perform \cite{Eberly:2023}. 
 Given a pure state (ensemble) decomposition (PSD) of a mixed state 
 $\rho = \sum_i p_i\ket{\psi_i}\bra{\psi_i}$ and an entanglement measure $E$ on pure states, one can form the average entanglement $E(\rho) = \sum_i p_i E(\ket{\psi_i})$. The entanglement measure on mixed states is then taken as the minimum of $E(\rho)$ over all possible PSDs $\{p_i,\ket{\psi_i}\}$ (the so called \tit{convex roof construction}). This final result is sometimes refered to as the entanglement of creation, which quantifies the resource required to create a given entangled state. Since $E(\rho)$ is a convex function one can apply Legendre transformations \cite{Eisert:2007} to transform the above difficult minimization into the ``minmax" problem \cite{Ghune:2007} 
 $E(\rho) = \underset{X}{\trm{max}}\,\underset{\psi}{\trm{min}}\{\Tr[X(\rho-\ket{\psi}\bra{\psi})] + E(\psi)\}$,
 where the exterior maximum is taken over all Hermitian matrices $X$. 
 The dimension of the exterior optimization over $X$ can be reduced to a rather small number by the symmetry of the state $\rho$, which greatly simplifies the numerical task, as was demonstrated in \cite{Ryu:2012}. 
 
\begin{figure}[h]
\begin{tabular}{cc}
\hspace{-0.25in}
\includegraphics[width=2.05in,height=1.5in]{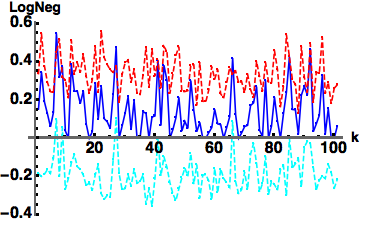} &
\includegraphics[width=2.05in,height=1.5in]{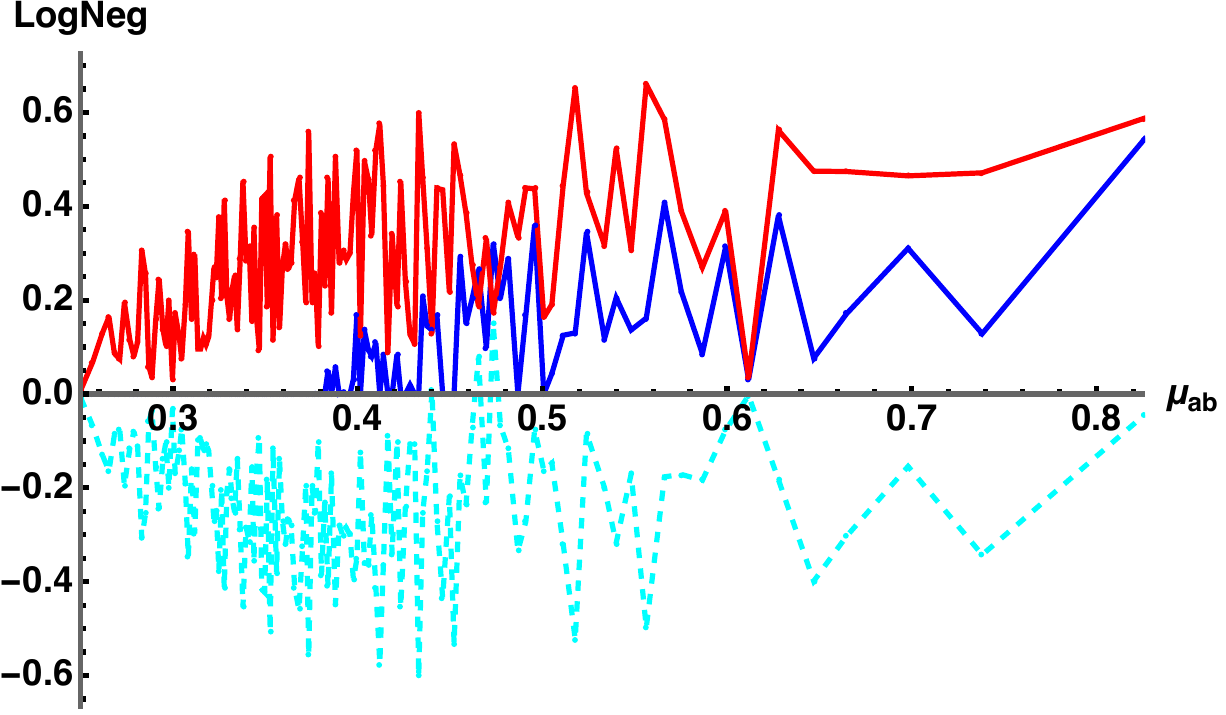}
\end{tabular}
\caption{$LN(\rho_{ab})$: 
(blue) exact numerical, 
(red) from \Eq{Neg:LogNeg:Approx:rhoab:line:3} 
(cyan) $LN_e-LN_a$,
for random $\rho_{ab}$ with 
(left) $M=40$,
(right) $M=1$ (two qubits).
}\label{fig:rho:random:M40}
\end{figure}

 In this work we are not interested in forming a proper entanglement measure, rather a witness, that can lower bound the exact Log Negativity.
Therefore, we will forgo the above computationally involved procedure and instead
 put forth (while somewhat simplistic, but computationally tractable)  an ansatz for a straightforward generalization of our pure state witness
 \bsub
 \bea{Neg:LogNeg:Approx:rhoab}
 \rho &=& \sum_{n,m}\sum_{n',m'} \rho_{nm,n'm'}\,\ket{n,m}\bra{n',m'}, \label{Neg:LogNeg:Approx:rhoab:line:1} \\
 \N_{a}&\to& \half \sum_{n,m\ne n} \left| \rho_{nn,mm} - \rho_{nm,mn}\right|, \qquad \label{Neg:LogNeg:Approx:rhoab:line:2} \\
 LN_{a} &=& \log_2\left( 1+2\,\N_{a}\right).\label{Neg:LogNeg:Approx:rhoab:line:3} 
 \eea
 \esub
 The main advantage of \Eq{Neg:LogNeg:Approx:rhoab:line:2} is that upon reduction to a pure state 
 $\rho_{nm,n'm'}\to c_{nm}\,c^*_{n'm'}$ the above formulas reduce to 
  \Eq{LogNeg:Approx:line:1} and  \Eq{LogNeg:Approx:line:2}.
  However, as \Fig{fig:rho:random:M40} shows, \Eq{Neg:LogNeg:Approx:rhoab:line:2} and \Eq{Neg:LogNeg:Approx:rhoab:line:3} now acts more as an approximate upper bound as the dimension $d$ increases (left), versus a desired lower bound, but typically at higher purity $\mu_{ab}\defn\Tr[\rho^2]$ values (right).
 Note that $LN_{a}(\rho_{ab})$ does a fairly good job of tracking the up and down random fluctuations of  
 $LN_{e}(\rho_{ab})$, but the former does not do a very good job when the latter is zero.
 On the negative side, \Eq{Neg:LogNeg:Approx:rhoab:line:2} does not capture separability when 
 $\rho_{ab} = \rho_a\otimes\rho_b$. 
 The issue of witnessing separability in general is a difficult, non-trivial problem \cite{Horodecki:1997}, 
 so it is not surprising that a simple generalization from a pure state witness to a  mixed state witness would not be valid.
 Nonetheless,  \Fig{fig:rho:random:M40} is intriguing for the relative tracking of 
 $LN_{a}(\rho_{ab})$ with $LN_{e}(\rho_{ab})$. $LN_{a}(\rho_{ab})$ 
 acts ``almost" as an upper bound  to  $LN_{e}(\rho_{ab})$, 
 (i.e. the cyan difference curve $LN_{e}(\rho_{ab})-LN_{a}(\rho_{ab})<0$),
 but there are places where it also acts as a lower bound, i.e.
 $LN_{e}(\rho_{ab})-LN_{a}(\rho_{ab}) > 0$.

 \subsection{$\boldsymbol{LN^{(Aprpox)}}$ on Werner states }
\begin{figure}[h]
\begin{tabular}{c}
\hspace{-0.25in}
\includegraphics[width=3.5in,height=2.25in]{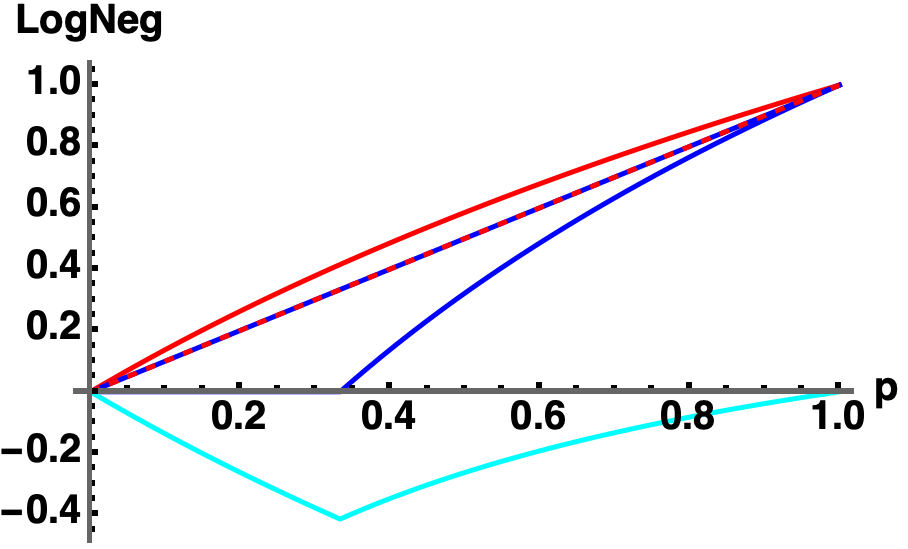} \\
\includegraphics[width=3.5in,height=2.25in]{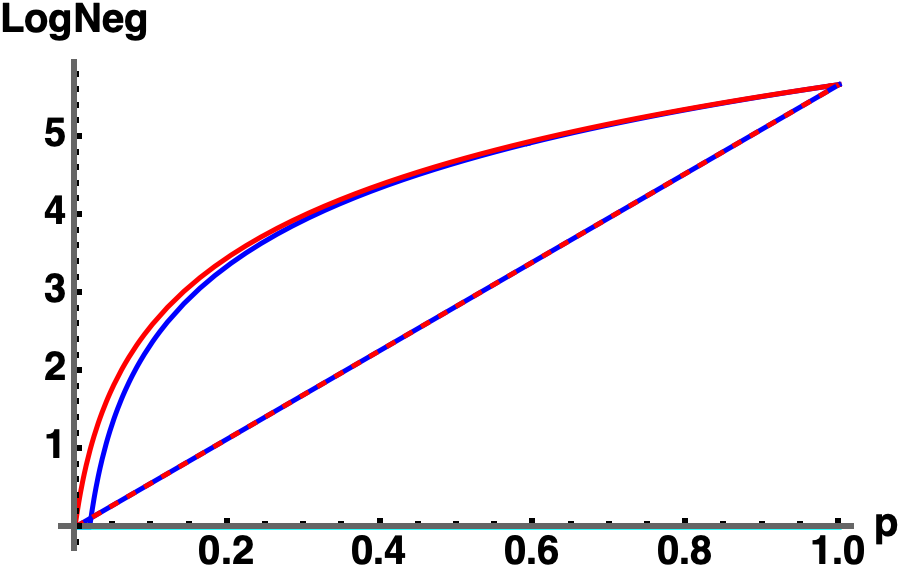}
\end{tabular}
\caption{$LN(\rho^{(W,\,d^2)}_{ab})$: for $d=(M+1)$ dimensional Werner states 
with probability  (to be in the pure state) $0\le p\le~1$:
(blue, solid) $LN_{e}(\rho_{ab})$, 
(red, solid) $LN_{a}(\rho_{ab})$ from \Eq{Neg:LogNeg:Approx:rhoab:line:3},
(cyan) $LN_{e}(\rho_{ab})-LN_{a}(\rho_{ab})$.
Average Log Legativity (ALN)
(blue, dashed) $LN^{avg}_{e}(\rho_{ab})$,
(red, dashed)  $LN^{avg}_{e}(\rho_{ab})$,
for uniformly random $\rho_{ab}$ with 
(top) $M=1$,
(bottom) $M=50$.
Note: $LN^{avg}_{e}(\rho_{ab})\equiv LN^{avg}_{e}(\rho_{ab})$ 
throughout all of $0\le p\le 1$.
(overlapping red-dashed and blue-dashed curves).
}\label{fig:rho:Werner:M1:M50}
\end{figure}

 Because they are analytically tractable,
 it is informative to examine the proposed  mixed state witness 
 \Eq{Neg:LogNeg:Approx:rhoab:line:3}
 for the case of Werner states of dimension $d^2$ 
 (where $d=(M+1)$), i.e
 $\rho^{(W,\,d^2)}_{ab}~=~p  \ket{\Psi_{Bell}}_{ab}\bra{\Psi_{Bell}} + \tfrac{(1-p)}{d^2} I_{d^2\times d^2}$ with 
 $\ket{\Psi_{Bell}}_{ab}~=~\frac{1}{\sqrt{d}}\sum_{n=0}^{d-1}\,\ket{nn}_{ab}$.
 (Note: $ \ket{\Psi_{Bell}}$ is a $d^2\times 1$ vector, so $\rho^{(W,\,d^2)}_{ab}$ is a $d^2\times d^2$ matrix).
 In \Fig{fig:rho:Werner:M1:M50} we show $LN_{e}$ and $LN_{a}$ for Werner states with 
 (top) $M=1$ (two qubits) and 
 (bottom) $M=50$ using the approximation for the Negativity given in \Eq{Neg:LogNeg:Approx:rhoab:line:3}.
Here we also introduce the Average Log Negativity (ALN) (for both exact and approximate $LN$) given by $LN^{avg}(\rho_{ab}) = \sum_i p_i\,LN(\rho^{(i)}_{ab})$ for mixed states of the form
$\rho_{ab} = \sum_i p_i\,\rho^{(i)}_{ab}$. 
It is straightforward to compute that for the Werner state 
the negative eigenvalues of the partial transpose are given by 
$\lambda_{-} = \tfrac{1}{d^2}\,\big(1-(d+1)p \big)$ for $\tfrac{1}{d+1}\le p\le 1$
with multiplicity
$\binom{d}{2} = \tfrac{d(d-1)}{2}$.
Therefore the negativity in this region is given by 
 $\N_e=\tfrac{1}{2} \tfrac{d-1}{d} \big((d+1)p-1\big)$ with 
 $LN_{e} = \log_2(1+2\,\N_{e})$
 yielding $LN_{e}(p=\frac{1}{d+1})=0$ and $LN_{e}(p=1)=\log_2(d)$.
Thus, $\rho^{(W,\,d^2)}_{ab}$ is entangled ($LN~>~0$) for $\tfrac{1}{(d+1)} <p\le 1$, 
and separable ($LN\le0$) for $0\le p\le\tfrac{1}{(d+1)}$. Finally, one can also show that the
purity $\mu(\rho)=\Tr[\rho^2]$ for the Werner states is given by
$\mu_{(W,\,d^2)}= \tfrac{1+(d^2-1)p}{d^2}$ corresponding to 
to a critical value $\mu^*_{(W,\,d^2)} = \tfrac{2}{d(d+1)}$ at the sudden death of entanglement
at $p= \tfrac{1}{(d+1)}$, where $\rho_{ab}$ becomes separable.

On the otherhand, for a given $d=M+1$ we have $\N_{a} = p\,(d-1)/2$ for the Werner states so that 
$LN_{a} = \log_2(1+(d-1)\,p)$, yielding again
$LN_{a}(p=1)=\log_2(d)$, but this time
 $LN_{a}=0$ at $p=0$ vs at $p=\tfrac{1}{d+1}$ as for $LN_{e}$.
Therefore, $LN_{a}(\rho_{ab})$ (red curve) never detects separability, and for Werner states, acts as a strict upper bound for $LN_{e}(\rho_{ab})$ (blue curve). Further, for Werner states  
$LN_{a}(\rho_{ab})\overset{d\gg 1}{\longrightarrow} LN_{e}(\rho_{ab})$ as demonstrated in the bottom plot in 
\Fig{fig:rho:Werner:M1:M50} with $M=50$. Note that in the limiting case of $d\to\infty$, the Werner state is always entangled, and never has a region of separability. 

The sudden death of entanglement  at $\mu^*_{(W,\,d^2)} = \tfrac{2}{d(d+1)}$ introduces a derivative discontinuity in  $LN_{e}$ that arises from the definition that the negativity is defined as non-zero only if some eigenvalues of the partial transpose of the density matrix are negative.  Equivalently, such a discontinuity could also be introduced for the Werner states by declaring that  $LN_{e}\ge 0$ \tit{iff} $\mu_{(W,\,d^2)}\ge\mu^*_{(W,\,d^2)}$. 
Our  definition of the continuous function $LN_{a}$ can never capture this derivative discontinuity. 
For Werner states one could attempt to capture this feature for any definition of a $LN_{a}$ by simply defining 
$LN_{a}\to 0$ for $\mu_{(W,\,d^2)}\le\mu^*_{(W,\,d^2)}$. 

A symmetrization of \Eq{Neg:LogNeg:Approx:rhoab:line:2} is given by
 \bea{LNA:sym:rho:ab}
 \N_{a}= \half \sum_{n,m\ne n} 
 &{}&\left| 
 \tfrac{1}{2}(\rho_{nn,mm} + \rho_{mm,nn})\right. \no
 &-&
 \left.
 \tfrac{1}{2}( \rho_{nm,mn} +  \rho_{mn,nm})
 \right|, \qquad 
 \eea
 which also reduces to \Eq{LogNeg:Approx:line:1} and  \Eq{LogNeg:Approx:line:2}
 for the case of pure states $\rho_{nm,n'm'}\to c_{nm}\,c^*_{n'm'}$ 
 \cite{LogNeg:rho:additonal:terms:comment}
 and has the additional favorable property that it \tit{does} detect separability $\rho = \sum_i p_i\,\rho_a^{(i)}\otimes\rho_b^{(i)}$
 \tit{if and only if, for each $i$, either $\rho_a^{(i)}$  or $\rho_a^{(i)}$ are $\in\mathbb{R}$}. This occurs because if
 $\rho_{nm,n'm'} = \sum_i p_i \, (\rho_a^{(i)})_{nn'}\, (\rho_b^{(i)})_{mm'}$, \Eq{LNA:sym:rho:ab} reduces to
 \bea{LNA:rhoab:on:mixed:states}
 \N_{a}\to \half \sum_{n,m\ne n} 
 &{}&\left| 
 \sum_i \tfrac{1}{2}\,p_i\,
 \Big( (\rho^{(i)}_a)_{nm}-(\rho^{*(i)}_a)_{nm} \Big)\right. \no
 &\times&
 \left.
 \Big( (\rho^{(i)}_a)_{nm}-(\rho^{*(i)}_a)_{nm} \Big)
 \right|, 
 \eea
where the terms inside the absolute value are proportional to the product of 
$\Im[(\rho^{(i)}_a)_{nm}]\,\Im[(\rho^{(i)}_b)_{nm}]$.

\Eq{LNA:sym:rho:ab} does detect a wide class of separable states, but of course, not all.
This is especially apparent since the application of  \Eq{LNA:sym:rho:ab} to the $\rho_{ab}^{(W,\,d^2)}$ produces exactly the same plots as shown in \Fig{fig:rho:Werner:M1:M50}.
We can understand this as follows. For the case of two qubits ($M=1$) we can write 
\bsub
\bea{rho:Werner:MS:form}
\hspace{-0.25in}
\rho_{ab}^{(W,\,d^2)} &=& \frac{1}{4}
\left( 
\mathbb{I} + p\,Z\otimes Z + p\,X\otimes X + p\,Y\otimes Y
\right)  \label{rho:Werner:MS:form:line:1} \\
&\equiv& \frac{(1-3p)}{4}\mathbb{I} + p\,\left( \mathbb{I} + Z\otimes Z\right) + p\,\left( \mathbb{I} + X\otimes X\right), \qquad\quad   \no
&{}& \hspace{0.6in} + \,p\,\left( \mathbb{I} - Y\otimes Y\right), \label{rho:Werner:MS:form:line:2} \\
&=&  \frac{(1-3p)}{4}\mathbb{I} + \frac{p}{2}\left(\ket{00}\bra{00}+\ket{11}\bra{11} \right), \no
&+& \frac{p}{2}\left(\ket{+-}\bra{+-}+\ket{-+}\bra{-+} \right) \no
&+& \frac{p}{2}\left(\ket{y_+y_-}\bra{y_+y_-}+\ket{y_-y_+}\bra{y_-y_+} \right). \label{rho:Werner:MS:form:line:3} 
\eea
\esub
While \Eq{rho:Werner:MS:form:line:1} 
is valid for all values of $0\le p\le 1$,
\Eq{rho:Werner:MS:form:line:2} and
\Eq{rho:Werner:MS:form:line:3} are only valid mixed state representations  
(since each of the four terms is diagonal in a separable basis, and positive semi-definite)
\tit{provided that} $p\le \tfrac{1}{3}$. Now the term
$\ket{y_+}\bra{y_+}\otimes\ket{y_-}\bra{y_-} = 
\tiny{
\tfrac{1}{2}\left(\begin{array}{cc}1 & -i \\i & 1 \end{array}\right)\otimes
\tfrac{1}{2}\left(\begin{array}{cc}1 & i \\-i & 1 \end{array}\right) 
} \equiv \rho_a^{(y)}\otimes\rho_b^{(y)}
$
involves density matrices $\rho_a^{(y)}$ and $\rho_b^{(y)}$ that are \tit{both} complex, 
for which our witness in \Eq{LNA:sym:rho:ab} cannot detect separability, since at least one of the  $\rho_a^{(y)}$ and $\rho_b^{(y)}$ must be real $\in\mathbb{R}$.

\subsection{Average $\boldsymbol{LN^{(Aprpox)}}$ as a lower bound for $\boldsymbol{LN_{e}}$}
\begin{figure}[h]
\begin{tabular}{c}
\hspace{-0.25in}
\includegraphics[width=3.5in,height=2.25in]{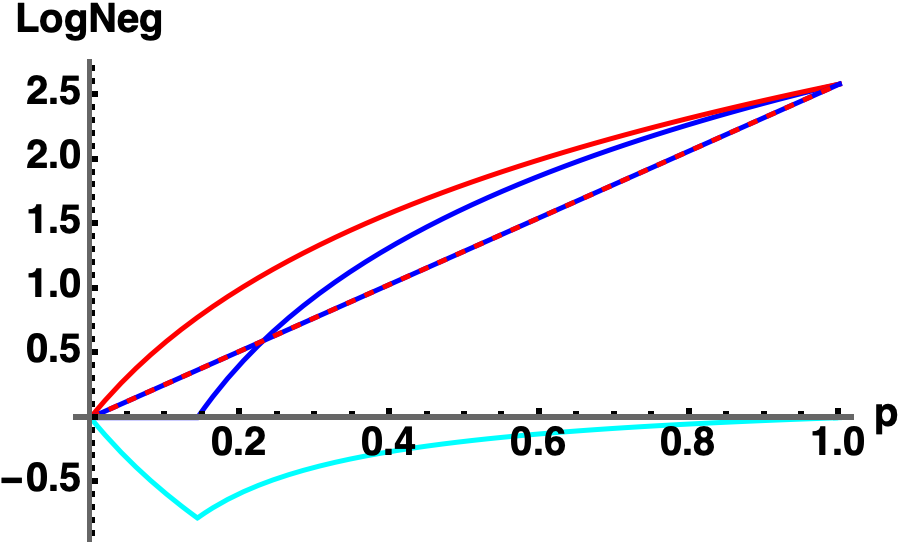} \\
\includegraphics[width=3.5in,height=2.25in]{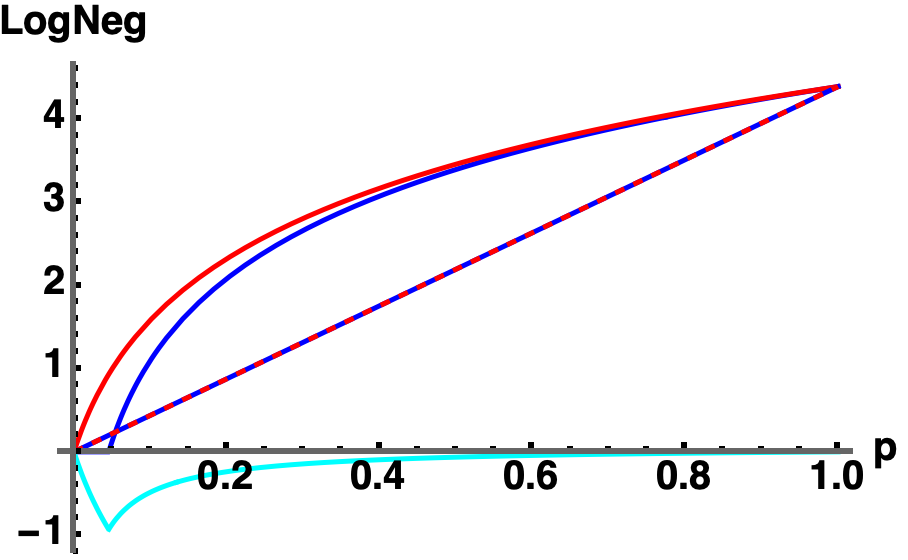}
\end{tabular}
\caption{$LN(\rho^{(W,\,d^2)}_{ab})$: for $d=(M+1)$ dimensional Werner states 
with probability  (to be in the pure state) $0\le p\le~1$:
(blue, solid) $LN_{e}(\rho_{ab})$, 
(red, solid) $LN_{a}(\rho_{ab})$ from \Eq{Neg:LogNeg:Approx:rhoab:line:3},
(cyan) $LN_{e}(\rho_{ab})-LN_{a}(\rho_{ab})$,
Average Log Negativity (ALN)
(blue, dashed) $LN^{avg}_{e}(\rho_{ab})$,
(red, dashed)  $LN^{avg}_{a}(\rho_{ab})$,
for uniformly random $\rho_{ab}$ with 
(top) $M=5$,
(bottom) $M=20$.
Note: $LN^{avg}_{e}(\rho_{ab})\equiv LN^{avg}_{e}(\rho_{ab})$ 
throughout all of $0\le p\le 1$.
(overlapping red-dashed and blue-dashed curves).
}\label{fig:rho:Werner:M5:M20}
\end{figure}

We note that for  Werner states , the average log negativity (ALN)
$LN^{avg}(\rho_{ab})=\sum_i p_i\,\rho^{(i)}_{ab}  \equiv \sum_i p_i\,LN(\rho^{(i)})$ 
for both the exact
$LN_{e}^{avg}(\rho_{ab})$, 
and the approximate version 
$LN_{a}^{avg}(\rho_{ab})$
(using \Eq{Neg:LogNeg:Approx:rhoab:line:3} on the righthand side), are  identically equal 
throughout all of $0\le p\le 1$, even in the separable region 
$0\le p\le\tfrac{1}{(d+1)}$ where $LN_{e}(\rho_{ab})=0$,
as shown as the overlapping dashed blue and red lines in \Fig{fig:rho:Werner:M5:M20}.
This is because  $LN_{e}=LN_{a}$ on the  $\rho^{(i)}$ components of the pure symmetric Bell state density matrix, and also on the maximally mixed state (yielding value $LN = 0$ on the latter).
In fact we see that the $LN_{e}$ 
is concave $LN_{e}(\rho)\ge LN_{a}^{avg}(\rho)$ in the region where entanglement is present, 
and is convex $LN_{e}(\rho)\le LN_{a}^{avg}(\rho)$ in the region where the state is separable.
On the other hand, our $LN_{a}(\rho)$ is concave throughout all of $0\le p \le 1$.
Thus, as  the dimension $d=M+1$ increases, 
$LN_{a}^{avg}(\rho)$ becomes an increasingly better lower bound for 
$LN_{e}(\rho)$ for a larger region of $p$, as the region of separability decreases,
as shown in \Fig{fig:rho:Werner:M5:M20} for $M=\{5,20\}$, 
and for $M=50$ in (bottom) \Fig{fig:rho:Werner:M1:M50}.

The above properties on Werner states
prompts us to compare $LN_{a}^{avg}(\rho_{ab})$ to $LN_{e}(\rho_{ab})$ for more general random density matrices written as a pure state ensemble decomposition (PSD), 
$\rho_{ab} = \sum_i p_i\,\ket{\psi_i}_{ab}\bra{\psi_i}$, which we consider (only)  
for the remainder of this section.
We also introduce an additional version of the average approximate Log Negativity given by
\be{mathcal:LN:avg}
\mathcal{LN}_{a}^{avg}(\rho_{ab})\defn \sum_i p_i\, LN_{a}\left(\ket{\psi_i}_{ab}\right),
\ee
which now uses our approximation of the  Log Negativity for pure states via \Eq{LogNeg:Approx:line:2} on the ensemble pure state components on the right hand side, which denoted as 
$LN_{a}\left(\ket{\psi_i}_{ab}\right)$.
This is to be distinguished from the notation $LN_a(\rho_{ab})$ 
which we will reserve to denote the Log Negativity using of our approximate mixed state formula \Eq{Neg:LogNeg:Approx:rhoab:line:3}, and 
$LN_a^{avg}(\rho_{ab}) \defn \sum_i p_i\,LN_a(\rho_{ab}^{(i)})$ which is the average approximate Log Negativity  that also uses \Eq{Neg:LogNeg:Approx:rhoab:line:3} on general mixed states
$\rho_{ab} = \sum_i p_i\,\rho_{ab}^{(i)}$.

\begin{figure}[h]
\begin{tabular}{c}
\hspace{-0.5in}
\includegraphics[width=3.25in,height=1.75in]{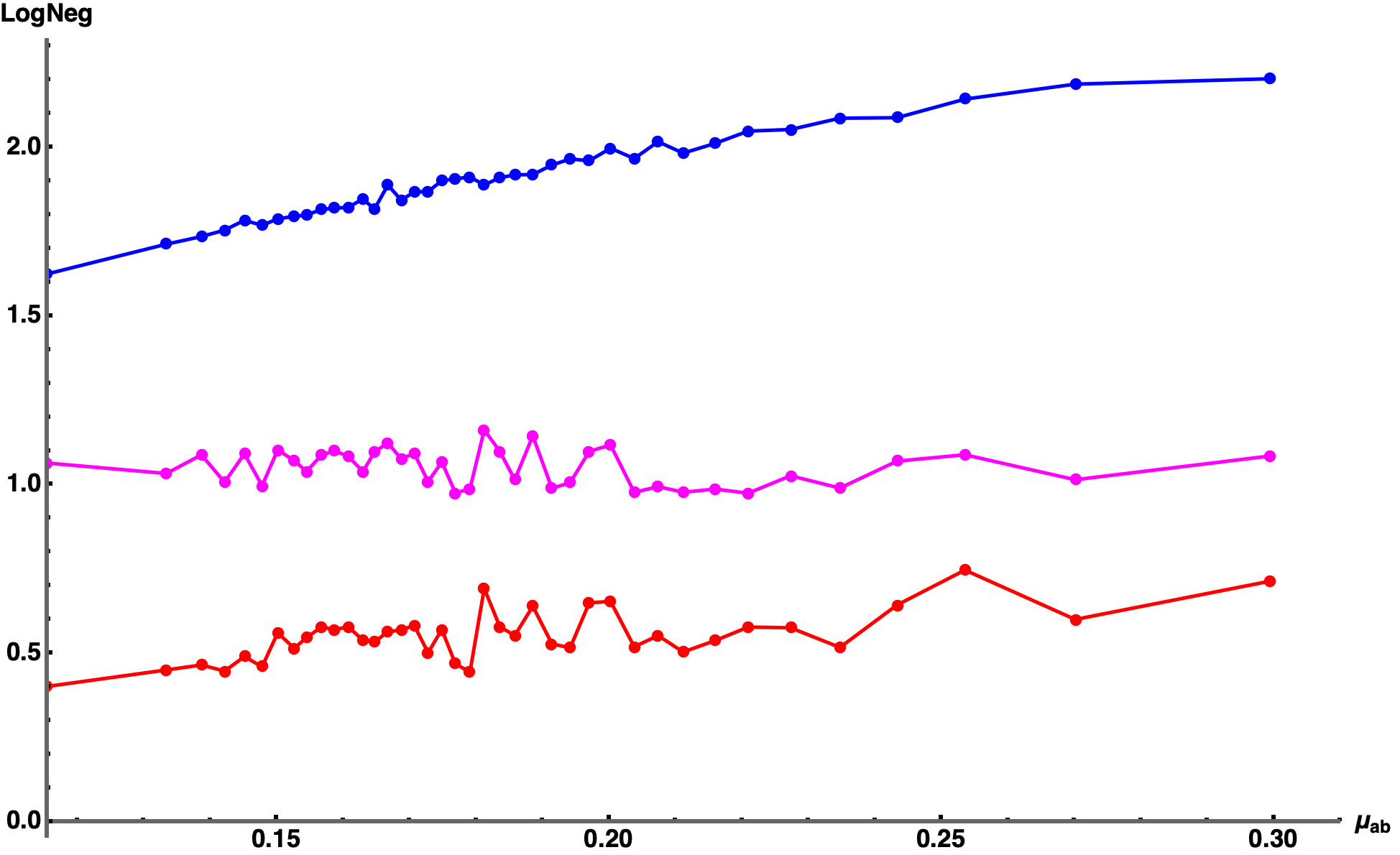} \\
\hspace{-0.5in}
\includegraphics[width=3.25in,height=1.75in]{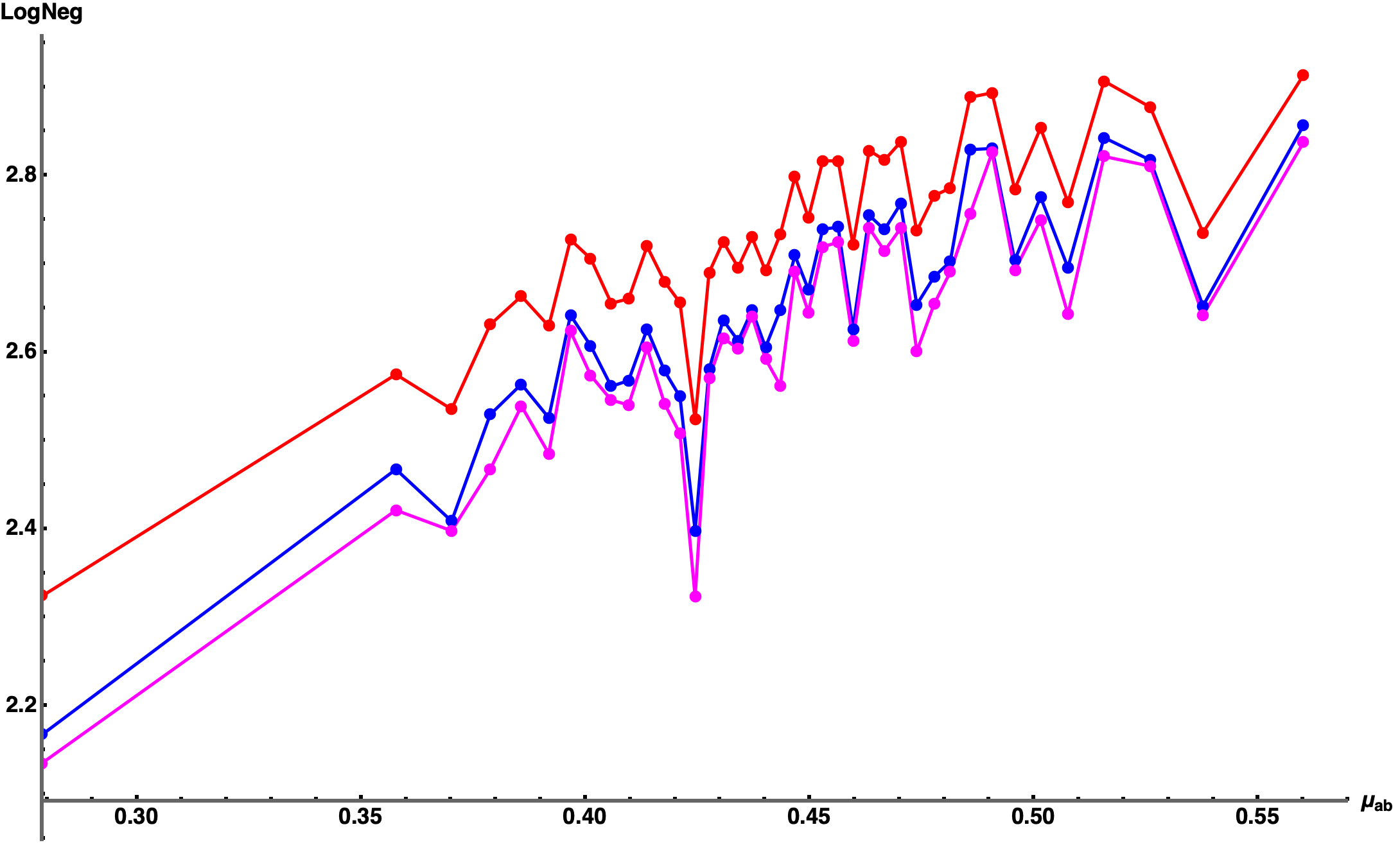} \\
\hspace{-0.5in}
\includegraphics[width=3.25in,height=1.75in]{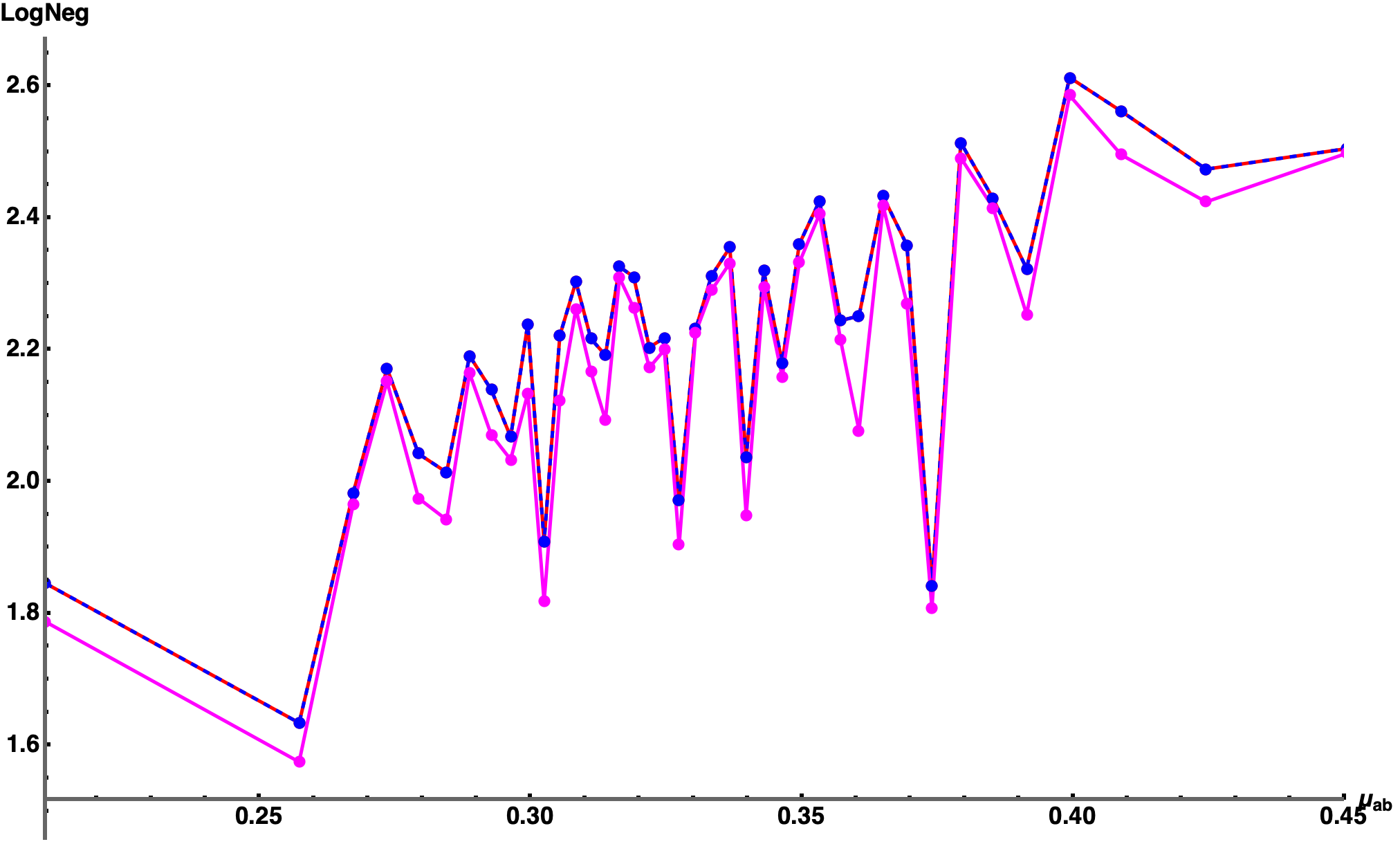}
\end{tabular}
\caption{$\mathcal{LN}_a^{avg}(\rho_{ab})$ and $LN_{a}(\rho_{ab})$  vs $LN_{e}(\rho_{ab})$ for 
PSDs  $\rho_{ab} = \sum_i p_i \ket{\psi_i}_{ab}\bra{\psi_i}$ with $M=10$, 
and $\ket{\psi_i} = \sum_{nm}^M C_{i,nm}\ket{n,m}_{ab}$,
plotted against purity 
$\mu_{ab}=\Tr[\rho^2_{ab}]$. 
Each $C_{i,nm}$ chosen as:
(top) a random complex matrix, $C_i\in\mathbb{C}$,
(middle) random Hermitian matrix $C_i = C^\dag_i$,
(bottom) random Orthogonal matrix $C_i = C^\trm{T}$.
Color legends:
(blue) $LN_{e}(\rho_{ab})$,
(magenta) $\mathcal{LN}_e^{avg}$,
(red) $LN_{a}(\rho_{ab})$.
The blue and red curves overlap in the bottom figure, $LN_a(\rho_{ab}) = LN_e(\rho_{ab})$.
%
}\label{fig:LNApprox:ALNApprox:vs:LNExact:M10:RC:U2:O2}
\end{figure}

The trend seen in the above Werner state discussion appears to hold in general for PSDs, 
namely that as the dimension $d=M+1$ 
of the of the pure state components $\ket{\psi_i}_{ab}$ increases, 
$\mathcal{LN}_{a}^{avg}(\rho_{ab})> LN_{a}(\rho_{ab})$ becomes an increasingly better proper lower bound of $LN_{e}(\rho_{ab})$,  as shown in 
\Fig{fig:LNApprox:ALNApprox:vs:LNExact:M10:RC:U2:O2}(top).
Here $M=10$ and $LN_{e}$ is given by the blue curve,
$LN_{a}$  by the red curve, and $\mathcal{LN}_{a}^{avg}$ by the magenta curve.
We are interested in the comparison of the red and magenta curves to the blue curve, 
based on how the $C_{i,nm}$ matrices are chosen in each pure state component of the PSD where 
 $\rho_{ab} = \sum_i p_i \ket{\psi_i}_{ab}\bra{\psi_i}$,  with 
$\ket{\psi_i} = \sum_{nm}^M C_{i,nm}\ket{n,m}_{ab}$.
%
%
In \Fig{fig:LNApprox:ALNApprox:vs:LNExact:M10:RC:U2:O2}(top) the $C_i\in\mathbb{C}$ are chosen as a random complex matrices. For this case we have
$LN_{a}(\rho_{ab})<\mathcal{LN}_{a}^{avg}(\rho_{ab})<LN_{e}(\rho_{ab})$ 
(red, magenta, blue curves)
as the dimension $d=M+1$ increases.
For \Fig{fig:LNApprox:ALNApprox:vs:LNExact:M10:RC:U2:O2}(middle) the $C_i = C^\dag_i$ are chosen as random Hermitian matrices. Here there is a dramatic change in structure. 
For this case we have
$\mathcal{LN}_{a}^{avg}<LN_{e}(\rho_{ab})<LN_{a}(\rho_{ab})$
(magenta, blue, red curves),
for which $LN_{a}(\rho_{ab})$ has switched to a proper upper bound of $LN_{e}(\rho_{ab})$,
and these inequalities hold all the down to $M=1$, the case of two qubits. Further, both 
$LN_{a}(\rho_{ab})$ and $\mathcal{LN}_{a}^{avg}(\rho_{ab})$ track the fluctuations of $LN_{e}(\rho_{ab})$ 
surprisingly, extremely well.
Finally, in \Fig{fig:LNApprox:ALNApprox:vs:LNExact:M10:RC:U2:O2}(bottom) the $C_i = C^\trm{T}$
are chosen as random Orthogonal matrices, and now, very surprisingly, we find that we have
$\mathcal{LN}_{a}^{avg}(\rho_{ab})<LN_{e}(\rho_{ab})=LN_{a}(\rho_{ab})$
(magenta,  blue overlapping red curves), 
namely that 
$LN_{a}(\rho_{ab})$ becomes exactly equal to $LN_{e}(\rho_{ab})$, while 
$\mathcal{LN}_{a}^{avg}(\rho_{ab})$ remains a lower bound 
(which appears somewhat tighter than in the previous Hermitian case).
Again, these inequalities hold all the down to $M=1$, the case of two qubits.

In general we have found that for $\rho$ written as a PSD,
$\mathcal{LN}_{a}^{avg}(\rho_{ab})$
based on our pure state negativity approximation given in  \Eq{LogNeg:Approx:line:2} 
does a better job as acting as a lower bound to 
$LN_{e}(\rho_{ab})$ than
 $LN_{a}(\rho_{ab})$
based on the mixed state negativity approximation given in \Eq{Neg:LogNeg:Approx:rhoab:line:3}.
%
As discussed in the Introduction \Eq{mixed:state:summary} 
and shown in \Fig{fig:LNApprox:ALNApprox:vs:LNExact:M10:RC:U2:O2},
we have numerically found that
\be{mixed:state:summary:again}
\mathcal{LN}_a^{avg}(\rho_{ab}) \leq LN_e(\rho_{ab}) \leq LN_a(\rho_{ab})
\ee
when the mixed state is written as a pure state decomposition (PSD) 
$\rho_{ab} = \sum_i p_i\,\ket{\psi_i}_{ab}\bra{\psi_i}$, 
 \tit{and in addition} the quantum amplitude matrix $C_i$
of each pure state component $\ket{\psi_i}_{ab} = \sum_{n m} C_{i,nm}\ket{n,m}_{ab}$ is  uniformly generated (over the Haar measure) as a positive Hermitian matrix, $C_i = C_i^\dag\ge0$.
Further, we find that the second inequality is surprisingly saturated, 
i.e. $LN_e(\rho_{ab}) = LN_a(\rho_{ab})$
 in \Eq{mixed:state:summary:again}, 
 if the uniformly randomly generated $C_i$ matrices above are real, i.e. $C_i=C_i^T$.
A discussion of the uniform generation of Hermitian matrices over the Haar measure
used in these numerical studies is given in \App{app:Haar:purity}.

\section{Summary and Conclusion}\label{sec:Conclusion}
In this work we presented numerical evidence for
a witness for  bipartite entanglement based solely on the coefficients $\{c_{nm}\}\to C$ 
of the pure state wavefunction $\ket{\psi}_{ab}=\sum_{nm} c_{nm} \ket{nm}_{ab}$, as opposed to the eigenvalues of the partial transpose of $\rho_{ab}=\ket{\psi}_{ab}\bra{\psi}$, 
as appropriate for the Log Negativity.
This is achieved by approximating the Negativity by diagonally dominant determinants of the coefficients as given in \Eq{LogNeg:Approx:line:1} . This $\N_{a}$ agrees exactly with the \tit{exact} Negativity  $\N_e$ (given by the standard definition obtained the sum of the absolute values of the negative eigenvalues of the partial transpose $\rho^\G_{ab}$) when 
$C= \{c_{nm}\}$ is a positive Hermitian matrix (i.e. $C$ is itself a non-unit trace density matrix). Interestingly, these include the cases when (i) $C$ is considered as a maximally mixed density matrix, then $\ket{\psi}_{ab}$ is the maximally entangled $d$-dimensional Bell state,  while (ii)  if $C$ is pure state density matrix,  then $\ket{\psi}_{ab}$ is separable, and (iii) symmetric superpositions of pure states with real coefficients. 
In these cases 
$\N_{a}(\ket{\psi}_{ab})=\N_{e}(\ket{\psi}_{ab})$. Of particular relevance is that for $C$ being a general complex matrix, we find that $\N_{a}(\ket{\psi}_{ab})$ acts as proper (but not tight) lower bound for 
$\N_{e}(\ket{\psi}_{ab})$, i.e. $\N_{a}(\ket{\psi}_{ab}) \le\N_{e}(\ket{\psi}_{ab})$.

In the second half of this work attempted to generalize our approximation of the negativity from pure states to mixed states via \Eq{Neg:LogNeg:Approx:rhoab:line:2}, which reduces to the
the  diagonally dominant determinants formula \Eq{LogNeg:Approx:line:1}  for pure states. We met with partial, yet still interesting, success. One of the key features of a Negativity based on  
\Eq{LNA:sym:rho:ab}, a symmetrized version of \Eq{LogNeg:Approx:line:1},
 is that it yields $LN_{a}(\rho_{ab})=0$ on separable states 
$\rho^{(sep)} = \sum_i p_i\,\rho_a^{(i)}\otimes \rho_b^{(i)}$ 
when either one of the component separable states 
$\rho_a^{(i)}$ or  $\rho_b^{(i)}$ is $\in\mathbb{R}$, real. 
While this of course does not capture all separable states (a non-trivial task) it does capture a wide range of physically relevant density matrices. 

In general we found that a $LN_{a}(\rho_{ab})$ based on \Eq{LNA:sym:rho:ab} acts more as an (undesired) upperbound to $LN_{e}(\rho_{ab})$, as demonstrated vividly on the analytically tractable $d$-dimensional Werner states. However, for Werner states we saw that 
$LN_{a}(\rho_{ab})\overset{d\gg 1}{\longrightarrow} LN_{e}(\rho_{ab})$ 
as the dimension $d$ grew large, and the corresponding region of separability 
grew smaller as $p_*=1/(d+1)$.

This led us to consider 
$\mathcal{LN}_{a}^{avg}(\rho_{ab}) \equiv \sum_i p_i LN_{a}(\ket{\psi_i}_{ab}\bra{\psi_i})$ for pure state ensemble decomposition (PSD) of the density matrices $\rho_{ab} = \sum_i p_i\,\ket{\psi_i}_{ab}\bra{\psi_i}$, now using our approximation of the pure state Negativity via \Eq{LogNeg:Approx:line:1}. 
Again, we saw that as the dimension $d$ of the 
pure state components $\ket{\psi_i}_{ab}$ increased, 
$\mathcal{LN}_{a}^{avg}(\rho_{ab})$ acted as a proper lower bound to 
$LN_{e}(\rho_{ab})$,  i.e. $\mathcal{LN}_{a}^{avg}(\rho_{ab})\overset{d\gg 1}{\le} LN_{e}(\rho_{ab})$. 
We speculate that this occurs for similar reasons as for the Werner states, namely that as the dimension $d$ increases, the region of separable states becomes vanishingly small \cite{Zyczkowski:1998,Zyczkowski_2ndEd:2020}, and simultaneously our Negativity formula \Eq{LogNeg:Approx:line:1}  is more accurate on pure states than  \Eq{Neg:LogNeg:Approx:rhoab:line:2} on mixed states, and hence the former does a better job as acting as a lower bound.
%

In summary, the main features of our proposed entanglement witness are 
(i) a simple formula for an approximate Negativity \Eq{LogNeg:Approx:line:1} on bipartite pure states directly in terms of the quantum amplitudes of the quantum states that is also exact for a certain class of physical relevant states (with the Log Negativity related to the the negative log of the purity of the quantum amplitude (matrix) when they have the properties of a normalized density matrix, \Eq{LN:Approx:cnm:random:U2:line:2}),
and
(ii) this Negativity formula does not require the need to numerically compute eigenvalues of the partial transpose of the density matrix.
An attempted generalization of this pure state Negativity formula for pure states to mixed states
yields that 
(iii)  in simulations on pure state decompositions (PSD) of density matrices
where the quantum amplitudes $C_i=\{c_{i, nm}\}$ of each of the pure state components 
$\ket{\psi_i}_{ab}$  act a positive Hermitian matrices,
$\mathcal{LN}_{a}(\rho_{ab})$ \Eq{mathcal:LN:avg}, based on the Log Negativity approximation for pure states \Eq{LogNeg:Approx:line:2}, yields a proper lower bound to $LN_{e}(\rho_{ab})$, and
(iv) on the PSD of density matrices in (iii), $LN_{a}(\rho_{ab})$ provides a proper upper bound to $LN_{e}(\rho_{ab})$ (with equality when  $C_i$ are Orthogonal matrices), and most interestingly,
both $\mathcal{LN}_{a}(\rho_{ab})$ and $LN_{a}(\rho_{ab})$ track, extremely well, the fluctuations in $LN_{e}(\rho_{ab})$ for uniformly randomly generated states.

As stated in the introduction, while the numerical computation of eigenvalues does not present a practical impediment to the calculation of entanglement measures, it is the surprising and unexpected relationship (and equality for certain physically relevant states, both pure and mixed) of the  entanglement witness we propose to the Log Negativity that was the impetus for this current investigation. It is our hope that this work might inspire further investigations into a more complete foundational derivation underpinning the results presented in these numerical investigations.

\clearpage
\newpage
\appendix
\section{Proof of \Eq{LN:Approx:cnm:random:U2}}\label{sec:app}
 From  \Eq{LogNeg:Approx:line:1} let us define $\tN_a$ 
 as the lower bound to the approximate Negativity $\N_{a}$ by
\bsub
\bea{LogNeg:Approx:app}
\N_{a}&=& \half\sum_n \sum_{m\ne n} \left| \trm{Det}\left(\begin{array}{cc} c_{nn} & c_{nm} \\c_{mn}& c_{mm}\end{array}\right)\right|,\qquad \label{LogNeg:Approx:line:1:app} \\
 &\ge& \left|\half\sum_n \sum_{m\ne n}  \trm{Det}\left(\begin{array}{cc} c_{nn} & c_{nm} \\c_{mn} &c_{mm}\end{array}\right)\right|, \no
 &\equiv&  \left|\tN^{(Approx)}\right|. \qquad  \label{LogNeg:Approx:line:2:app}
\eea
\esub
Now 
\bea{tildeN:calc:C}
\tN_a &=& \half\sum_n \sum_{m\ne n} \left( c_{nn} \, c_{mm} -c_{nm}\,c_{mn} \right), \no
&=& 
\half\sum_n c_{nn} \Big[\big(\sum_m c_{mm}\big) -c_{nn}\Big],\no
&-& \half\Big[\big(\sum_n \sum_m c_{nm}  c_{mn}\big) -\sum_n c^2_{nn}\Big],\no
&=& \half(\sum_n c_{nn}\big)(\sum_m c_{mm}\big) - \half\sum_n \sum_m c_{nm}  c_{mn}, \no
&=& \half\left(\left( \Tr[C ]\right)^2 - \Tr[C^2 ]\right).
\eea
Recall that $\ket{\psi} = \sum_{n,m} c_{nm}\ket{n,m}_{ab} $
with $C=\{c_{nm}\}$ a complex matrix subject only to the constraint
$1=\IP{\psi}{\psi}=\Tr[C C^\dag] = 1$.
Thus let us define another arbitrary complex matrix $A$ via
\be{A:defn}
C = \frac{A}{\sqrt{\Tr[A A^\dag]}} \quad\Rightarrow\quad \Tr[C C^\dag] = 1,
\ee
noting that $A A^\dag>0$ is a positive matrix.
Thus \Eq{tildeN:calc:C} becomes
\be{tildeN:A}
\tN_a= \half\frac{ \left(\Tr[A ]\right)^2 - \Tr[A^2 ]}{\Tr[A A^\dag ]}.
\ee

Now let us impose some additional conditions on $A$ and hence $C$.

\flushleft{\bf{Case 1}:} $A = A^\dag$ is Hermitian.
\bsub
\bea{tildeN:A:Hermitian}
\tN_a^{A=A^\dag}&=& \half\frac{ \left(\Tr[A ]\right)^2 - \Tr[A^2 ]}{\Tr[A^2 ]}, \\
&=& \half \left( \frac{\left(\Tr[A ]\right)^2}{\Tr[A^2 ]} -  1\right), \\
LN_{a} &=& \log_2\left( \frac{\left(\Tr[A ]\right)^2}{\Tr[A^2 ]}  \right).
\eea
\esub

\bsub
\flushleft{\bf{Case 2}:} $A = A^\dag$ is Hermitian and $\Tr[A]=1$.
Note that  $\Tr[A]=1$ does \tit{not} imply that $A$ is positive, since $A$ could have
negative real values on the diagonal such that $\Tr[A]=1$.
\bea{tildeN:A:Hermitian:TrAeq1}
\tN_a^{A=A^\dag, \Tr[A]=1} &=& \half \left( \frac{1}{\Tr[A^2 ]} -  1\right), \\
LN_{a} &=& \log_2\left( \frac{1}{\Tr[A^2 ]}  \right).
\eea
\esub
\medskip

\flushleft{\bf{Case 3}:} $A\equiv\rho = \rho^\dag $ is Hermitian, positive $\rho>0$ and $\Tr[\rho]\ne1$, i.e.
$A=\rho$ is an unnormalized density matrix.
Then
\bsub
\bea{tildeN:A:Hermitian:TrAneq1}
\tN_a^{A\to\rho=\rho^\dag, \rho>0, \Tr[\rho]\ne1} &=& \half \left( \frac{\left(\Tr[\rho]\right)^2}{\Tr[\rho^2 ]} -  1\right), \label{tildeN:A:Hermitian:TrAneq1:line1}\\
LN_{a} &=& \log_2\left( \frac{\left(\Tr[\rho ]\right)^2}{\Tr[\rho^2 ]}  \right). \label{tildeN:A:Hermitian:TrAneq1:line2}
\eea
\esub
\medskip
It is in this latter Case 3 that we find numerically that $\tN^{(Approx)}=\N_{a}=\N^{(Exact)}$.
\Eq{tildeN:A:Hermitian:TrAneq1} is \Eq{LN:Approx:cnm:random:U2:line:1}.
\newline

Lastly, we consider the case 
\flushleft{\bf{Case 4}:} $A\equiv\rho = \rho^\dag $ is Hermitian, positive $\rho>0$ \tit{and} $\Tr[\rho]=1$,  i.e.
$A=\rho$ is a proper density matrix itself.
Then
\bsub
\bea{tildeN:A:Hermitian:TrAeq1}
\tN_a^{A\to\rho=\rho^\dag, \rho>0, \Tr[\rho]=1} &=& \half \left( \frac{1}{\Tr[\rho^2 ]} -  1\right), \\
&\equiv& \half \left( \frac{1}{\mu(\rho)} -  1\right), \\
LN_{a} &=& \log_2\left( \frac{1}{\Tr[\rho^2 ]}  \right),\\
C &=& \frac{\rho}{\sqrt{\Tr[\rho^2]}}.
\eea
\esub
where $1/D \le\mu(\rho)\equiv \Tr[\rho^2]\le 1$ is the \tit{purity} of the density matrix $\rho$ of dimension $D$.
\medskip

\clearpage
\newpage
\section{Random unitary matrices, and sampling of density matrices over purity ranges}\label{app:Haar:purity}
\begin{lstlisting}[mathescape,language=Mathematica,
caption={Mathematica code to generate random unitary $U$},label=Mathematica:Code:2,frame=single]
URandom[n_] := 
Module[{Z, Q, R, diagR, $\Lambda$}, 
 RG:=RandomVariate[NormalDistribution[]];
 
 Z = Table[$\tfrac{1}{\sqrt{2}}$(RG + I RG), {n}, {n})];
 {Q,R}=QRDecomposition[Z];
 
 (* Note: Z=Q.R=(Q $\Lambda$).($\Lambda^{-1}$R).*)
 (*Make R (hence Q) unique by forcing R*)
 (*to have positive diagonal entries*)
 
 diagR=Diagonal[R];
 
 (* diagonal matrix $\tfrac{R_{ii}}{|R_{ii}|}$*)
 (* Note: $\Lambda^{-1}$R makes diagonal entries *)
 (* of $R$ to be $|R_{ii}|$*)

$\Lambda$=DiagonalMatrix[diagR/Abs[diagR]]//Chop;
  
 (* return unique unitary matrix *)
 Q = Q . $\Lambda$ // Chop
]
\end{lstlisting}
\begin{lstlisting}[mathescape,language=Mathematica,
caption={Mathematica code to generate random $\rho$},label=Mathematica:Code:3,frame=single]
$\rho\trm{R}$andom[n_] := 
Module[{U,$\rho$d}, 
 RI:=RandomInteger[{1,n}];
 
 (*$\rho$d = $\rho_{diagonal}$: take a random row of  U*)
 U=URandom[n];
 $\rho$d = Abs[ U[[RI]] ]$^2$ // Chop;
 
 (*form $\rho = U . \rho\trm{d} . U^\dag $*)
  U=URandom[n];
  
 (*return random density matrix $\rho$ *)
 $\rho$ = U.$\rho$d.U$^\dag$ // Chop
]
\end{lstlisting}
The above are the \tit{Mathematica} codes were used to generate uniformly random unitary (and orthogonal) matrices (via the Haar measure), and subsequently uniformly random density matrices, based on the article by Mezzadri \cite{Mezzadri:2007}. The central idea behind these codes is that the space of invertible $N\times N$ complex matrices $Z=\{z_{jk}\}$ (the Ginibre ensemble) have matrix elements that are independent identically distributed (i.i.d.) standard normal complex random variables with probability distribution $p(z_{jk})=\tfrac{1}{\pi} e^{|z_{jk}|^2}$.
The joint probability distribution for the matrix elements (also statistically independent) is given by
$P(Z) = \tfrac{1}{\pi^{N^2}} \prod_{j,k=1}^N e^{-|z_{jk}|^2} = 
 \tfrac{1}{\pi^{N^2}} \exp\left[\sum_{j,k=1}^N e^{-|z_{ij}|^2}\right]$ $= \tfrac{1}{\pi^{N^2}} \exp\left(-\Tr[Z^\dag\,Z]\right)$.
$P(Z)$ is normalized to unity via $\int_{\mathbb{C}^{N^2}} P(Z)\, dZ=1$ where
$dZ=\prod_{j,k=1}^N dx_{jk}\,dy_{jk}$ and $z_{jk} = x_{jk} + i\,y_{jk}$. The integration measure on the Ginibre ensemble $\mathbb{C}^{N\times N}\cong \mathbb{C}^{N^2}$ is $d\mu_G(Z) = P(Z)\,dZ$ which can be thought of as an infinitesimal volume in $\mathbb{C}^{N^2}$. 
The crucial point is that $d\mu_G(Z)$ is invariant under left and right multiplications of $Z$ by arbitrary unitary matrices i.e. $d\mu_G(U\,Z)=d\mu_G(Z\,V)=d\mu_G(Z)$ for  $U, V\in U(N)$. The proof follows trivially from the property of the trace in the definition of $P(Z)$ since 
$\Tr[ (U\,Z)^\dag\,(U\,Z)]= \Tr[ (Z\,V)^\dag\,(Z\,V)]=\Tr[Z^\dag\,Z]$, and hence $P(U\,Z)=P(Z\,V)=P(Z)$.
This Haar measure is the matrix analogue of a uniform probability distribution in one dimension 
$p(\theta) = \tfrac{1}{2\,\pi}$.

The above algorithm to generate random unitary matrices distributed with the Haar measure uses the $QR$ decomposition of $Z$ (vs the less stable, at higher dimensions, Gram-Schmidt orthonormalization routine).  
Here $Q$ is a unitary matrix and $R$ is an upper-triangular matrix.
One caveat exits though. The $QR$ decomposition is not unique, since if $Z=Q\,R$ then so is
$Z=Q'\,R'\equiv (Q\,\Lambda) \, (\Lambda^{-1}\,R)$, with $\Lambda$ a diagonal matrix so that $Q'$ and $R'$ are again unitary and upper-triangular matrices. To make the decomposition unique, the solution (see Mezzadri \cite{Mezzadri:2007} for details) is to choose $\Lambda_{ii} = R_{ii}/|R_{ii}|$ where $R_{ii}$ are the diagonal elements of $R$. Thus, the diagonal matrix elements of $R'=(\Lambda^{-1}\,R)$ are real and strictly positive, which renders 
the matrix $Q'=Q\,\Lambda$ unique and uniformly distributed with the Haar measure. This is the procedure implemented in the above Mathematica codes. 

Given a code to generate random unitary matrices $U$, one can subsequently generate uniform density matrices
$\rho$ (positive Hermitian matrices of unit trace)
 by the simple decomposition $\rho = U\,\rho_{diagonal}\, U^\dag >0$. To generate the diagonal density matrix 
$\rho_{diagonal}$ (an element of the Weyl chamber, section 8.5 of \cite{Zyczkowski_2ndEd:2020}) such that
$\sum_{i=1}^N (\rho_{diagonal})_{ii}=1$, one simply generates another random unitary $U'$ and takes the absolute square of the a random row (or column) as the the diagonal entries. This is implemented in the second code listed above. These codes are easily implementable in other commonly used coding languages, such as Python.

While the codes above generate uniformly (Haar) distributed density matrices $\rho$, the purity of such generated matrices are distributed non-linearly. This is because the Haar measure enforces linear constraint $\Tr[\rho]=1$ on the eigenvalues of $\rho$, the purity is given by the non-linear (quadratic) constraint $\tfrac{1}{N}\le\mu(\rho)= \Tr[\rho^2]\le 1$.
In general a Haar distributed $\rho$ tends to sigmoid shaped distribution in purity, with sampling favoring lower values of the purity \cite{Zyczkowski:1998}.
Recently, the authors have developed an algorithm for sampling density matrices for fixed values of the purity \cite{Alsing_DDM:2022}, although here adopt a numerically simpler procedure.

Any particular individual state 
is a set of measure zero in the Weyl chamber,
for example, $\rho=\ket{Bell^{d^2}}_{ab}\bra{Bell^{d^2}}$, with $\ket{Bell^{d^2}}_{ab}$ the $d^2$ maximally entangled Bell state of unit purity. 
Thus to uniformly sample states about a particular state of interest, 
it advantageous to construct uniform 
deviations about the chosen state, with some adjustable parameter.
To ensure that we were sampling over the full range of purity values 
in this work we generated matrices  of the form 
$\rho_{ab} = \rho'_{ab}/\Tr[\rho'_{ab}]$ 
where (for example) $\rho'_{ab}=\ket{Bell^{d^2}}_{ab}\bra{Bell^{d^2}} + \eta\,\sigma_{ab}$ 
with $\sigma$ a Haar uniformly distributed density matrix, and $\eta>0\in\mathbb{R}$ a real scaling factor. By empirically adjusting $\eta$ for a given dimension $d=M+1$ (with $M$ the maximum Fock state in either subspace) we could adjust the mean value of the purity
$\mu(\rho_{ab})$ sampled according to the Haar measure.

\begin{figure}[!ht]
\begin{tabular}{cc}
\includegraphics[width=1.75in,height=1.25in]{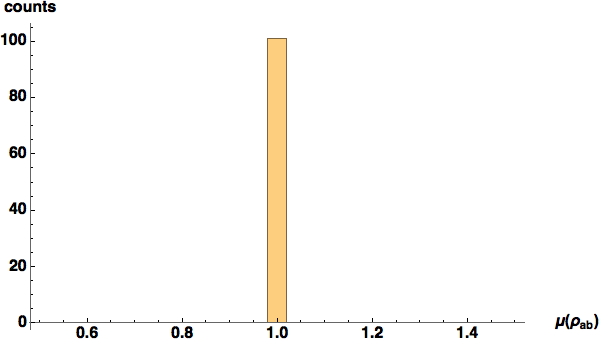} &
\includegraphics[width=1.75in,height=1.25in]{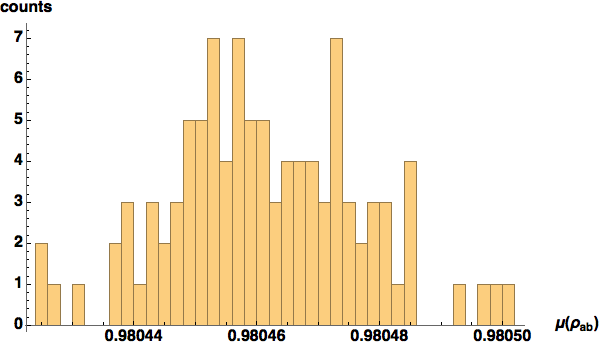} \\
\includegraphics[width=1.75in,height=1.25in]{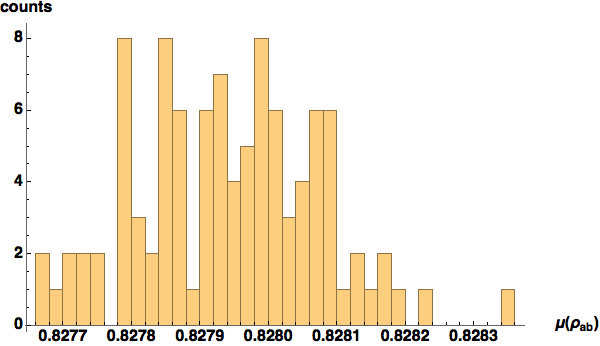} &
\includegraphics[width=1.75in,height=1.25in]{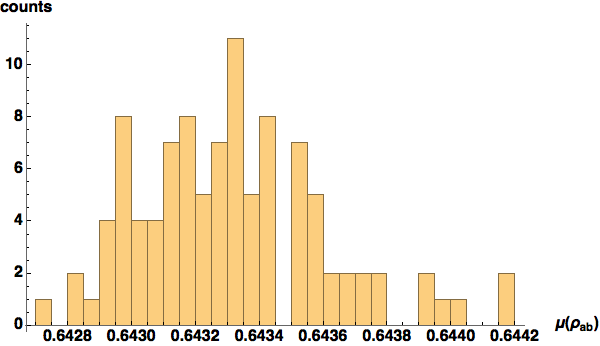} 
\end{tabular}
\caption{Histograms of purities $\mu(\rho_{ab})$ of Haar distributed density matrices of the form
$\rho_{ab} = \rho'_{ab}/\Tr[\rho'_{ab}]$ 
with $\rho'_{ab}=\ket{Bell^{d^2}}_{ab}\bra{Bell^{d^2}} + \eta\,\sigma_{ab}$ for
(top left) $\eta= 0$,
(top right)  $\eta= 0.01$,
(bottom left)  $\eta= 0.1$,
(bottom right)   $\eta= 0.25$
}\label{fig:histograms:M10:eta:0:p01:p1:p25}
\end{figure}

\begin{figure}[!h]
\begin{tabular}{cc}
\includegraphics[width=1.75in,height=1.25in]{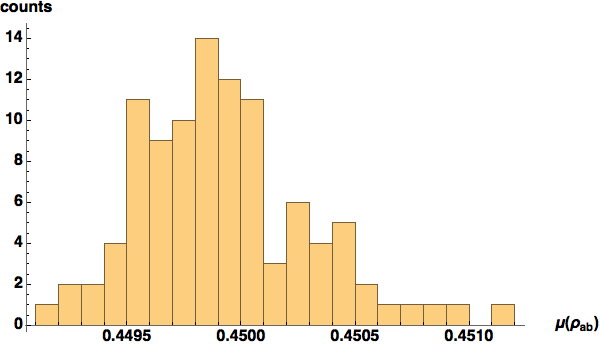} &
\includegraphics[width=1.75in,height=1.25in]{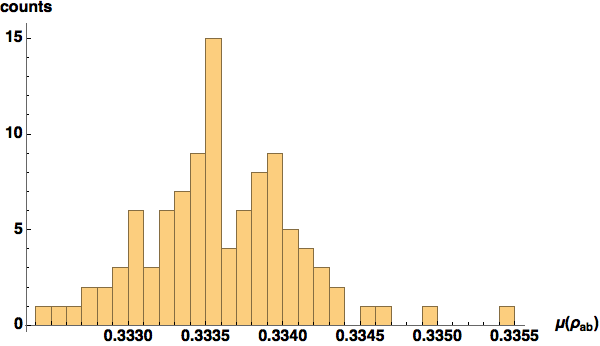} \\
\includegraphics[width=1.75in,height=1.25in]{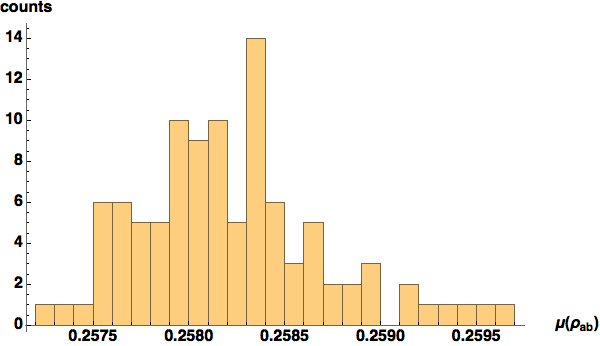} &
\includegraphics[width=1.75in,height=1.25in]{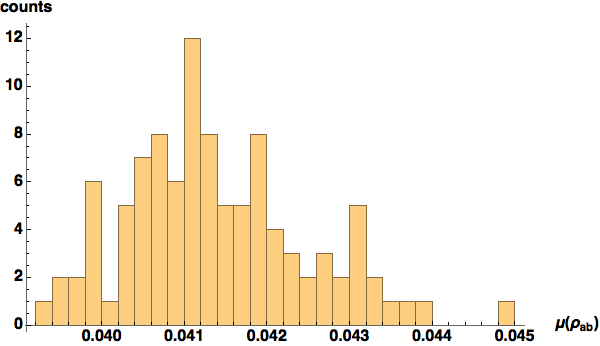} \\
\includegraphics[width=1.75in,height=1.25in]{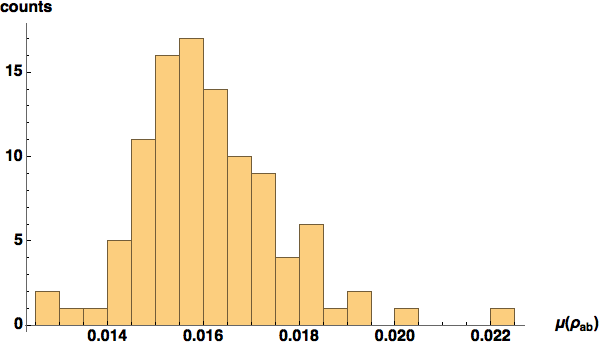} &
\includegraphics[width=1.75in,height=1.25in]{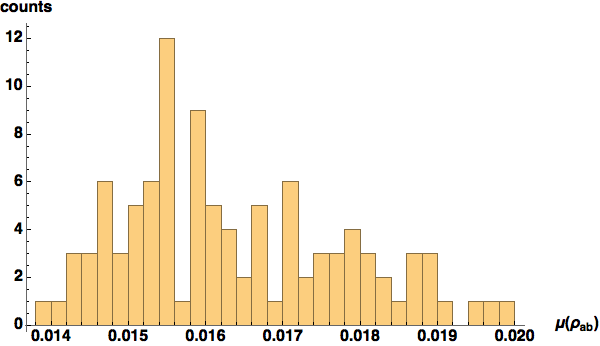} 
\end{tabular}
\caption{Same as \Fig{fig:histograms:M10:eta:0:p01:p1:p25}, now with
(top left)          $\eta= 0.50$,
(top right)        $\eta= 0.75$,
(middle left)     $\eta= 1$,
(middle right)   $\eta= 5$
(bottom left)     $\eta= 100$,
(bottom right)   $\eta= 1000$
}\label{fig:histograms:M10:eta:p5:p75:1:5:100:1000}
\end{figure}

In \Fig{fig:histograms:M10:eta:0:p01:p1:p25} and \Fig{fig:histograms:M10:eta:p5:p75:1:5:100:1000}
we show histograms of $100$ samples/plot of $\mu(\rho_{ab})$ for $d^2\times d^2$ Haar distributed 
density matrices $\rho_{ab}$ with
$\eta=\{0, 0.01, 0.10, 0.25, 0.50, 0.75, 1.0, 5, 100, 1000\}$ for $M=10, \, d=M+1=11,\, d^2=121$.
Fine tuning $\eta$ (for each fixed $M$) allows one to narrow in on a mean value $\EV{\mu(\rho_{ab})}$.
This procedure was use to generate plots such as 
\Fig{fig:rho:eq:rho0:plus:eta:sigma:M20} and \Fig{fig:rho:eq:MMS:plus:eta:sigma:M20}.

\clearpage
\newpage
\begin{acknowledgments}
Any opinions, findings and conclusions or recommendations
expressed in this material are those of the author(s) and do not
necessarily reflect the views of the authors home instituion.
\end{acknowledgments}



\begin{thebibliography}{23}%
\makeatletter
\providecommand \@ifxundefined [1]{%
 \@ifx{#1\undefined}
}%
\providecommand \@ifnum [1]{%
 \ifnum #1\expandafter \@firstoftwo
 \else \expandafter \@secondoftwo
 \fi
}%
\providecommand \@ifx [1]{%
 \ifx #1\expandafter \@firstoftwo
 \else \expandafter \@secondoftwo
 \fi
}%
\providecommand \natexlab [1]{#1}%
\providecommand \enquote  [1]{``#1''}%
\providecommand \bibnamefont  [1]{#1}%
\providecommand \bibfnamefont [1]{#1}%
\providecommand \citenamefont [1]{#1}%
\providecommand \href@noop [0]{\@secondoftwo}%
\providecommand \href [0]{\begingroup \@sanitize@url \@href}%
\providecommand \@href[1]{\@@startlink{#1}\@@href}%
\providecommand \@@href[1]{\endgroup#1\@@endlink}%
\providecommand \@sanitize@url [0]{\catcode `\\12\catcode `\$12\catcode
  `\&12\catcode `\#12\catcode `\^12\catcode `\_12\catcode `\%12\relax}%
\providecommand \@@startlink[1]{}%
\providecommand \@@endlink[0]{}%
\providecommand \url  [0]{\begingroup\@sanitize@url \@url }%
\providecommand \@url [1]{\endgroup\@href {#1}{\urlprefix }}%
\providecommand \urlprefix  [0]{URL }%
\providecommand \Eprint [0]{\href }%
\providecommand \doibase [0]{http://dx.doi.org/}%
\providecommand \selectlanguage [0]{\@gobble}%
\providecommand \bibinfo  [0]{\@secondoftwo}%
\providecommand \bibfield  [0]{\@secondoftwo}%
\providecommand \translation [1]{[#1]}%
\providecommand \BibitemOpen [0]{}%
\providecommand \bibitemStop [0]{}%
\providecommand \bibitemNoStop [0]{.\EOS\space}%
\providecommand \EOS [0]{\spacefactor3000\relax}%
\providecommand \BibitemShut  [1]{\csname bibitem#1\endcsname}%
\let\auto@bib@innerbib\@empty
\bibitem [{\citenamefont {Toth}\ and\ \citenamefont
  {Apellaniz}(2014)}]{Toth:2014}%
  \BibitemOpen
  \bibfield  {author} {\bibinfo {author} {\bibfnamefont {G.}~\bibnamefont
  {Toth}}\ and\ \bibinfo {author} {\bibfnamefont {I.}~\bibnamefont
  {Apellaniz}},\ }\bibfield  {title} {\enquote {\bibinfo {title} {Quantum
  metrology from quantum information science perspective},}\ }\href@noop {}
  {\bibfield  {journal} {\bibinfo  {journal} {J. Phys. A: Mathematical and
  Theoretical}\ }\textbf {\bibinfo {volume} {47}},\ \bibinfo {pages} {424006}
  (\bibinfo {year} {2014})}\BibitemShut {NoStop}%
\bibitem [{\citenamefont {of~teh DOE Quantum Internet
  Blueprint~Workshop}(5-6Feb2020)}]{QWkshpReport:2014}%
  \BibitemOpen
  \bibfield  {author} {\bibinfo {author} {\bibfnamefont {Report}\ \bibnamefont
  {of~teh DOE Quantum Internet Blueprint~Workshop}},\ }\href@noop {} {\enquote
  {\bibinfo {title} {From long-distance entanglement to building a nationwide
  quantum network},}\ }\bibinfo {howpublished}
  {\url{https://www.energy.gov/sites/prod/files/2020/07/f76/QuantumWkshpRpt20FINAL_Nav_0.pdf}}
  (\bibinfo {year} {5-6Feb2020})\BibitemShut {NoStop}%
\bibitem [{\citenamefont {Nielsen}\ and\ \citenamefont
  {Chuang}(2000)}]{NC:2000}%
  \BibitemOpen
  \bibfield  {author} {\bibinfo {author} {\bibfnamefont {M.A.}\ \bibnamefont
  {Nielsen}}\ and\ \bibinfo {author} {\bibfnamefont {I.L.}\ \bibnamefont
  {Chuang}},\ }\href@noop {} {\emph {\bibinfo {title} {Quantum Computation and
  Quantum Information}}}\ (\bibinfo  {publisher} {Cambridge University Press,
  Cambridge, UK},\ \bibinfo {year} {2000})\BibitemShut {NoStop}%
\bibitem [{\citenamefont {Schr{\"{o}}dinger}(1935)}]{Schrodinger:1935}%
  \BibitemOpen
  \bibfield  {author} {\bibinfo {author} {\bibfnamefont {E.}~\bibnamefont
  {Schr{\"{o}}dinger}},\ }\bibfield  {title} {\enquote {\bibinfo {title}
  {Discussion of probability relations between separated systems},}\
  }\href@noop {} {\bibfield  {journal} {\bibinfo  {journal} {Mathematical
  Proceedings of the Cambridge Philosophical Society}\ }\textbf {\bibinfo
  {volume} {31}},\ \bibinfo {pages} {555} (\bibinfo {year} {1935})}\BibitemShut
  {NoStop}%
\bibitem [{\citenamefont {Schr{\"{o}}dinger}(1936)}]{Schrodinger:1936}%
  \BibitemOpen
  \bibfield  {author} {\bibinfo {author} {\bibfnamefont {E.}~\bibnamefont
  {Schr{\"{o}}dinger}},\ }\bibfield  {title} {\enquote {\bibinfo {title}
  {Probability relations between separated systems},}\ }\href@noop {}
  {\bibfield  {journal} {\bibinfo  {journal} {Mathematical Proceedings of the
  Cambridge Philosophical Society}\ }\textbf {\bibinfo {volume} {32}},\
  \bibinfo {pages} {446} (\bibinfo {year} {1936})}\BibitemShut {NoStop}%
\bibitem [{\citenamefont {Einstein}\ \emph {et~al.}(1935)\citenamefont
  {Einstein}, \citenamefont {Podolsky},\ and\ \citenamefont
  {Rosen}}]{EPR:1935}%
  \BibitemOpen
  \bibfield  {author} {\bibinfo {author} {\bibfnamefont {A.}~\bibnamefont
  {Einstein}}, \bibinfo {author} {\bibfnamefont {B.}~\bibnamefont {Podolsky}},
  \ and\ \bibinfo {author} {\bibfnamefont {N.}~\bibnamefont {Rosen}},\
  }\bibfield  {title} {\enquote {\bibinfo {title} {Can quantum-mechanical
  description of physical reality be considered complete},}\ }\href@noop {}
  {\bibfield  {journal} {\bibinfo  {journal} {Phys. Rev.}\ }\textbf {\bibinfo
  {volume} {47}},\ \bibinfo {pages} {777} (\bibinfo {year} {1935})}\BibitemShut
  {NoStop}%
\bibitem [{\citenamefont {Swingle}(2017)}]{Swingle:2017}%
  \BibitemOpen
  \bibfield  {author} {\bibinfo {author} {\bibfnamefont {B.}~\bibnamefont
  {Swingle}},\ }\bibfield  {title} {\enquote {\bibinfo {title} {Spacetime from
  entanglement},}\ }\href@noop {} {\bibfield  {journal} {\bibinfo  {journal}
  {Annual Review of Cond. Matt. Phys.}\ }\textbf {\bibinfo {volume} {9}},\
  \bibinfo {pages} {345} (\bibinfo {year} {2017})}\BibitemShut {NoStop}%
\bibitem [{\citenamefont {Horodecki}\ \emph {et~al.}(2009)\citenamefont
  {Horodecki}, \citenamefont {Horodecki}, \citenamefont {M.Horodecki},\ and\
  \citenamefont {Horodecki}}]{Horodecki4:2009}%
  \BibitemOpen
  \bibfield  {author} {\bibinfo {author} {\bibfnamefont {R.}~\bibnamefont
  {Horodecki}}, \bibinfo {author} {\bibfnamefont {P.}~\bibnamefont
  {Horodecki}}, \bibinfo {author} {\bibnamefont {M.Horodecki}}, \ and\ \bibinfo
  {author} {\bibfnamefont {K.}~\bibnamefont {Horodecki}},\ }\bibfield  {title}
  {\enquote {\bibinfo {title} {Quantum entanglement},}\ }\href@noop {}
  {\bibfield  {journal} {\bibinfo  {journal} {Reviews of Modern Physics}\
  }\textbf {\bibinfo {volume} {81}},\ \bibinfo {pages} {865} (\bibinfo {year}
  {2009})}\BibitemShut {NoStop}%
\bibitem [{\citenamefont {Bengtsson}\ and\ \citenamefont
  {Zyczkowski}(2020)}]{Zyczkowski_2ndEd:2020}%
  \BibitemOpen
  \bibfield  {author} {\bibinfo {author} {\bibfnamefont {I.}~\bibnamefont
  {Bengtsson}}\ and\ \bibinfo {author} {\bibfnamefont {K.}~\bibnamefont
  {Zyczkowski}},\ }\href@noop {} {\emph {\bibinfo {title} {The Geometry of
  Quantum States, 2nd Ed.}}}\ (\bibinfo  {publisher} {Cambridge University
  Press,Cambridge},\ \bibinfo {year} {2020})\BibitemShut {NoStop}%
\bibitem [{\citenamefont {Wootters}(1998)}]{Wootters:1998}%
  \BibitemOpen
  \bibfield  {author} {\bibinfo {author} {\bibfnamefont {W.K.}\ \bibnamefont
  {Wootters}},\ }\bibfield  {title} {\enquote {\bibinfo {title} {Entanglement
  of formation of an arbitrary state of two qubits},}\ }\href@noop {}
  {\bibfield  {journal} {\bibinfo  {journal} {Phys. Rev. Lett.}\ }\textbf
  {\bibinfo {volume} {80}},\ \bibinfo {pages} {2245} (\bibinfo {year}
  {1998})}\BibitemShut {NoStop}%
\bibitem [{\citenamefont {Peres}(1996)}]{Peres:1996}%
  \BibitemOpen
  \bibfield  {author} {\bibinfo {author} {\bibfnamefont {A.}~\bibnamefont
  {Peres}},\ }\bibfield  {title} {\enquote {\bibinfo {title} {Separability
  criterion for density matrices},}\ }\href@noop {} {\bibfield  {journal}
  {\bibinfo  {journal} {Phys. Rev. Lett.}\ }\textbf {\bibinfo {volume} {77}},\
  \bibinfo {pages} {1413} (\bibinfo {year} {1996})}\BibitemShut {NoStop}%
\bibitem [{\citenamefont {Horodecki}(1997)}]{Horodecki:1997}%
  \BibitemOpen
  \bibfield  {author} {\bibinfo {author} {\bibfnamefont {P.}~\bibnamefont
  {Horodecki}},\ }\bibfield  {title} {\enquote {\bibinfo {title} {Separability
  criterion and inseparable mixed states with positive partial
  transposition},}\ }\href@noop {} {\bibfield  {journal} {\bibinfo  {journal}
  {Phys. Lett. A}\ }\textbf {\bibinfo {volume} {232}},\ \bibinfo {pages} {333}
  (\bibinfo {year} {1997})}\BibitemShut {NoStop}%
\bibitem [{\citenamefont {Agarwal}(2013)}]{Agarwal:2013}%
  \BibitemOpen
  \bibfield  {author} {\bibinfo {author} {\bibfnamefont {G.~S.}\ \bibnamefont
  {Agarwal}},\ }\href@noop {} {\emph {\bibinfo {title} {Quantum Optics}}}\
  (\bibinfo  {publisher} {Cambridge University Press},\ \bibinfo {address}
  {Cambridge},\ \bibinfo {year} {2013})\BibitemShut {NoStop}%
\bibitem [{mey()}]{meyer:wallach:comment}%
  \BibitemOpen
  \href@noop {} {\bibinfo  {journal} {The closest related antisymmetric
  structure we have found is the norm-squared of the wedge product
  $D(u,v)=\sum_{n<m} |u_n\,v_n - u_m\,v_n|^2$ used in Meyer and Wallach's
  multiparticle global entanglement monotone $Q$ \cite{Meyer_Wallach:2002}
  involving the pure states $u=\sum_n u_n \ket{n}_a$ and $v=\sum_m v_m
  \ket{m}_b$}\ }\BibitemShut {NoStop}%
\bibitem [{\citenamefont {Meyer}\ and\ \citenamefont
  {Wallach}(2002)}]{Meyer_Wallach:2002}%
  \BibitemOpen
\bibfield  {journal} {  }\bibfield  {author} {\bibinfo {author} {\bibfnamefont
  {D.A.}\ \bibnamefont {Meyer}}\ and\ \bibinfo {author} {\bibfnamefont {N.R.}\
  \bibnamefont {Wallach}},\ }\bibfield  {title} {\enquote {\bibinfo {title}
  {Global entanglement in multi-particle systems},}\ }\href@noop {} {\bibfield
  {journal} {\bibinfo  {journal} {J. Math. Phys.}\ }\textbf {\bibinfo {volume}
  {43}},\ \bibinfo {pages} {4273} (\bibinfo {year} {2002})}\BibitemShut
  {NoStop}%
\bibitem [{\citenamefont {Xie}\ \emph {et~al.}(2023)\citenamefont {Xie},
  \citenamefont {Younis},\ and\ \citenamefont {Eberly}}]{Eberly:2023}%
  \BibitemOpen
  \bibfield  {author} {\bibinfo {author} {\bibfnamefont {S.}~\bibnamefont
  {Xie}}, \bibinfo {author} {\bibfnamefont {D.}~\bibnamefont {Younis}}, \ and\
  \bibinfo {author} {\bibfnamefont {J.~H.}\ \bibnamefont {Eberly}},\ }\bibfield
   {title} {\enquote {\bibinfo {title} {Evidence for unexpected robustness of
  multipartite entanglement against sudden death from spontaneous emission},}\
  }\href@noop {} {\bibfield  {journal} {\bibinfo  {journal} {Phys. Rev.
  Research}\ }\textbf {\bibinfo {volume} {5}},\ \bibinfo {pages} {L032015}
  (\bibinfo {year} {2023})}\BibitemShut {NoStop}%
\bibitem [{\citenamefont {Eisert}\ \emph {et~al.}(2007)\citenamefont {Eisert},
  \citenamefont {Brand{\~{a}}o},\ and\ \citenamefont
  {Audenaert}}]{Eisert:2007}%
  \BibitemOpen
  \bibfield  {author} {\bibinfo {author} {\bibfnamefont {J.}~\bibnamefont
  {Eisert}}, \bibinfo {author} {\bibfnamefont {F.~G.}\ \bibnamefont
  {Brand{\~{a}}o}}, \ and\ \bibinfo {author} {\bibfnamefont {K.~M.}\
  \bibnamefont {Audenaert}},\ }\bibfield  {title} {\enquote {\bibinfo {title}
  {Quantitative entanglement witnesses},}\ }\href@noop {} {\bibfield  {journal}
  {\bibinfo  {journal} {New J. Phys.}\ }\textbf {\bibinfo {volume} {9}},\
  \bibinfo {pages} {46} (\bibinfo {year} {2007})}\BibitemShut {NoStop}%
\bibitem [{\citenamefont {Gh{\"{u}}hne}\ \emph {et~al.}(2007)\citenamefont
  {Gh{\"{u}}hne}, \citenamefont {Reimpell},\ and\ \citenamefont
  {Werner}}]{Ghune:2007}%
  \BibitemOpen
  \bibfield  {author} {\bibinfo {author} {\bibfnamefont {O.}~\bibnamefont
  {Gh{\"{u}}hne}}, \bibinfo {author} {\bibfnamefont {M.}~\bibnamefont
  {Reimpell}}, \ and\ \bibinfo {author} {\bibfnamefont {R.~F.}\ \bibnamefont
  {Werner}},\ }\bibfield  {title} {\enquote {\bibinfo {title} {Estimating
  entanglement measures in experiments},}\ }\href@noop {} {\bibfield  {journal}
  {\bibinfo  {journal} {Phys. Rev. Lett.}\ }\textbf {\bibinfo {volume} {98}},\
  \bibinfo {pages} {110502} (\bibinfo {year} {2007})}\BibitemShut {NoStop}%
\bibitem [{\citenamefont {Ryu}\ \emph {et~al.}(2012)\citenamefont {Ryu},
  \citenamefont {Lee},\ and\ \citenamefont {Sim}}]{Ryu:2012}%
  \BibitemOpen
  \bibfield  {author} {\bibinfo {author} {\bibfnamefont {S.}~\bibnamefont
  {Ryu}}, \bibinfo {author} {\bibfnamefont {S.~S.~B.}\ \bibnamefont {Lee}}, \
  and\ \bibinfo {author} {\bibfnamefont {H.-S.}\ \bibnamefont {Sim}},\
  }\bibfield  {title} {\enquote {\bibinfo {title} {Minimax optimization of
  entanglement witness operator for the quantification of three-qubit
  mixed-state entanglement},}\ }\href@noop {} {\bibfield  {journal} {\bibinfo
  {journal} {Phys. Rev. Lett.}\ }\textbf {\bibinfo {volume} {86}},\ \bibinfo
  {pages} {042324} (\bibinfo {year} {2012})}\BibitemShut {NoStop}%
\bibitem [{Log()}]{LogNeg:rho:additonal:terms:comment}%
  \BibitemOpen
  \href@noop {} {\bibinfo  {journal} {Note that one could also add terms such
  as $\half\,\sum_{n}\sum_{n\ne m} \left( \rho_{nn,nn}\,\rho_{mm,mm}-
  \rho_{nn,mm}\,\rho_{mm,nn}\right)$ which reduce to zero on the pure state
  case $\rho_{nm,n'm'}\to$ $c_{nm}\,c^*_{n'm'}$. However, we have not found
  such additional terms useful}\ }\BibitemShut {NoStop}%
\bibitem [{\citenamefont {Mezzadri}(2007)}]{Mezzadri:2007}%
  \BibitemOpen
\bibfield  {journal} {  }\bibfield  {author} {\bibinfo {author} {\bibfnamefont
  {F.}~\bibnamefont {Mezzadri}},\ }\bibfield  {title} {\enquote {\bibinfo
  {title} {How to generate random matrices from the classical compact
  groups},}\ }\href@noop {} {\bibfield  {journal} {\bibinfo  {journal} {Notices
  of the AMS}\ }\textbf {\bibinfo {volume} {54(5)}},\ \bibinfo {pages} {592}
  (\bibinfo {year} {2007})}\BibitemShut {NoStop}%
\bibitem [{\citenamefont {Zyczkowski}\ \emph {et~al.}(1998)\citenamefont
  {Zyczkowski}, \citenamefont {Horodecki}, \citenamefont {Sanpera},\ and\
  \citenamefont {Lewenstein}}]{Zyczkowski:1998}%
  \BibitemOpen
  \bibfield  {author} {\bibinfo {author} {\bibfnamefont {K.}~\bibnamefont
  {Zyczkowski}}, \bibinfo {author} {\bibfnamefont {P.}~\bibnamefont
  {Horodecki}}, \bibinfo {author} {\bibfnamefont {A.}~\bibnamefont {Sanpera}},
  \ and\ \bibinfo {author} {\bibfnamefont {M.}~\bibnamefont {Lewenstein}},\
  }\bibfield  {title} {\enquote {\bibinfo {title} {Volume of the set of
  separable states},}\ }\href@noop {} {\bibfield  {journal} {\bibinfo
  {journal} {Phys. Rev. A}\ }\textbf {\bibinfo {volume} {58(2)}},\ \bibinfo
  {pages} {883} (\bibinfo {year} {1998})}\BibitemShut {NoStop}%
\bibitem [{\citenamefont {Alsing}\ \emph {et~al.}(2022)\citenamefont {Alsing},
  \citenamefont {Tison}, \citenamefont {Schneeloch}, \citenamefont
  {Birrittella},\ and\ \citenamefont {Fanto}}]{Alsing_DDM:2022}%
  \BibitemOpen
  \bibfield  {author} {\bibinfo {author} {\bibfnamefont {P.~M.}\ \bibnamefont
  {Alsing}}, \bibinfo {author} {\bibfnamefont {C.~C.}\ \bibnamefont {Tison}},
  \bibinfo {author} {\bibfnamefont {J.~S.}\ \bibnamefont {Schneeloch}},
  \bibinfo {author} {\bibfnamefont {R.~J.}\ \bibnamefont {Birrittella}}, \ and\
  \bibinfo {author} {\bibfnamefont {M.~L.}\ \bibnamefont {Fanto}},\ }\bibfield
  {title} {\enquote {\bibinfo {title} {The distribution of density matrices at
  fixed purity for arbitrary dimensions},}\ }\href@noop {} {\bibfield
  {journal} {\bibinfo  {journal} {Phys. Rev. Research}\ }\textbf {\bibinfo
  {volume} {4}},\ \bibinfo {pages} {043114} (\bibinfo {year}
  {2022})}\BibitemShut {NoStop}%
\end{thebibliography}
\providecommand{\noopsort}[1]{}\providecommand{\singleletter}[1]{#1}%
%

%

\end{document}